\documentclass[onecolumn,preprintnumbers,nofootinbib,prd]{revtex4}
\usepackage{sidecap}
\usepackage{wrapfig}

\usepackage{mathrsfs}

\usepackage{graphicx}
\usepackage{amssymb}
\usepackage{color}
\usepackage{enumerate}
\usepackage{amsmath}
\usepackage{slashed}
\usepackage{comment}
\usepackage{cancel}
\def\beq{\begin{equation}}
\def\eeq{\end{equation}}
\def\beqn{\begin{eqnarray}}
\def\eeqn{\end{eqnarray}}

\def\con{\color{black}} %n for nera

\renewcommand{\texttt}{{}}
\newcommand{\be}{\begin{eqnarray}}
\newcommand{\ee}{\end{eqnarray}}

\begin{document}

\title{
{\bf{
Finite quantum gravity in dS and AdS spacetimes}}}

\author{Alexey S. Koshelev}
\email{alexey@ubi.pt}

\affiliation{Departamento de F\'{i}sica, Centro de Matem\'atica e Aplica\c{c}\~oes
	(CMA-UBI), Universidade da Beira Interior, 6200 Covilh\~a, Portugal}

\affiliation{Theoretische Natuurkunde, Vrije Universiteit Brussel and \\ The International
	Solvay Institutes, Pleinlaan 2, B-1050 Brussels, Belgium}

\author{K. Sravan Kumar}
\email{sravan@ubi.pt}
\affiliation{Departamento de F\'{i}sica, Centro de Matem\'atica e Aplica\c{c}\~oes
	(CMA-UBI), Universidade da Beira Interior, 6200 Covilh\~a, Portugal}
	\affiliation{Department of Physics, Southern University of Science and Technology, Shenzhen 518055, China}

\author{Leonardo Modesto}
\email{lmodesto@sustc.edu.cn} %, lmodesto1905@icloud.com}
\affiliation{Department of Physics, Southern University of Science and Technology, Shenzhen 518055, China}

\author{Les\l{}aw Rachwa\l{}}
\email{grzerach@gmail.com}
\affiliation{Department of Physics, Southern University of Science and Technology, Shenzhen 518055, China}
\affiliation{Instituto de F\'{i}sica, Universidade de Bras\'{i}lia, 70910-900, Bras\'{i}lia, DF, Brazil}

%{\color{red}AFFILIATIONS }

%%%%%%%%%%%%%%%%%%%%%%%%%%%%%%
%%%%%%%%%%%%%%%%%%%%%%%%%%%%%%

%\date{\small\today}

\begin{abstract} \noindent
We hereby study the properties of a large class of weakly nonlocal gravitational theories around the (anti-) de Sitter spacetime background. In particular we explicitly prove that the kinetic operator for the graviton field has the same structure as the one in Einstein-Hilbert theory around any maximally symmetric spacetime.
Therefore, the perturbative spectrum is the same of standard general relativity, while the propagator on any maximally symmetric spacetime is a mere generalization of the one from Einstein's gravity derived and extensively studied in several previous papers. At quantum level the range of theories here presented is superrenormalizable or finite when proper (non affecting the propagator) terms cubic or higher in curvatures are added. Finally, it is proven that for a large class of nonlocal theories, which in their actions do involve neither the Weyl nor the Riemann tensor, the theory is classically equivalent to the Einstein-Hilbert one with cosmological constant by means of a metric field redefinition at any perturbative order.
\end{abstract}

\maketitle

%\begin{keyword}
%perturbative
%quantum gravity %\sep nonlocal field theory
%\PACS{04.60.-m} %, 11.10.Lm}
%\end{keyword}

%\tableofcontents

\section{Introduction}

In previous studies it has been extensively shown that a class of weakly nonlocal
theories of gravity is unitary (ghost-free) and perturbatively superrenormalizable or finite in the framework of quantum field theory
\cite{Kuzmin, Krasnikov, Tombo, Khoury, modesto, modestoLeslaw, Briscese:2013lna, Mtheory,M3,M4,NLsugra}. These works mostly concentrated on the perturbative theory around the flat Minkowski spacetime.
The very foundations of the theory are the following:
(i) general covariance;
(ii) weak nonlocality (or quasipolynomiality) \cite{Efimov};
(iii) unitarity (ghost freedom);
(iv) superrenormalizability or finiteness at quantum level\footnote{We here would like to point out that in this paper all the results are proved in dimensional regularization scheme (DIMREG). This is to clarify the difference with the Wilsonian point of view and Functional Renormalization Group approach where the cut-off is taken seriously. We will use the standard DIMREG scheme adopted in QED, in QCD, and all the Standard Model of Particle Physics. However, our results are of course independent on the regularization scheme. We can, for example, use the cut-off regularization scheme with Pauli-Villars fields as it is done in QED and we end up with exactly the same results found in DIMREG. Finally, the readers interested in the analysis of nonlocal theories carried out in the cut-off regularization scheme, we refer to \cite{entanglement}.}.
The new class of generally covariant theories differs from Einstein's gravity because of the weak nonlocality, which makes possible to achieve unitarity and superrenormalizability at the same time, and at any order in the perturbative loop expansion.
Nevertheless, the theory is not unique and all the freedom is mainly encoded
in one, two, or three form factors (entire functions) with very specific asymptotic limits in the ultraviolet (UV) and infrared (IR) regimes in order to have a well defined quantum field theory.

We here study the same range of weakly nonlocal theories around maximally symmetric
spacetimes (MSS) applying exactly the same logic so successfully implemented for theories around
the Minkowski vacuum. In particular, we show that the kinetic operator for the gravitational fluctuations {\bf h} resumes exactly the Einstein-Hilbert one up to some multiplicative factors. For this achievement, we explicitly show the results for the expansion of the action at the second order in {\bf h} around a MSS.
Therefore, all the results concerning the propagator on (anti-) de Sitter [(A)dS] spaces for the Einstein-Hilbert action
can be exported and applied to the quasipolynomial theories too.
In particular, for one out of the two classes of theories, which we extensively study in this paper, we prove by the means of a field redefinition that at perturbative level, but to all perturbative orders in the field redefinition, the nonlocal action is classically equivalent to the Einstein-Hilbert one in the presence of a cosmological constant. The proof is based on a field redefinition theorem %\cite{Anselmi:2006yh, Anselmi:2002ge}
that was already applied in \cite{amplitudes} to the theory around the Minkowski vacuum.

There are several good theoretical as well as observational reasons to study
the class of gravitational theories around MSS and not only around flat spacetime vacuum.
Primarily, the true gravitational vacuum in quantum field theory is not precisely located
as suggested by the cosmological constant problem. This has to do in other disguises with the gravitational effect of the zero modes of the simple quantized harmonic oscillator. (The last one works as a toy model for any perturbative quantum field theory (QFT), when it is treated as a theory of free propagating excitations.) Therefore, the flat Minkowski spacetime may not be the correct gravitational vacuum and in some theories this state may even decay (via the spontaneous production of ghosts) like in higher derivative models of gravity. It may happen that the perturbative calculus around such false vacua is very fast divergent and not reliable due to the presence of different type of instabilities such as ghosts (negative norm states) or tachyons (negative mass states). A rescue could be to look for another vacuum state and to study quantum perturbations around the new vacuum.
The MSS are the only other spacetimes where the number of local generators is not in conflict with the one of the Poincar\'{e} group for the flat spacetime. On MSS the spacetime symmetries are as rich as when on flat spacetime; hence MSS is potentially another good vacuum state. (We remind that the vacuum state is a state of quite high symmetry.)
On such spacetime we do not violate homogeneity neither isotropy and the group of symmetries is only changed from $SO(1,3)$ into $SO(2,2)$ in the case of AdS in $D=4$ spacetime dimensions. In the case of dS the group of isometries remains the same. These new different vacua may be perturbatively unreachable from the original one, therefore, by studying quantum theories around (A)dS spacetimes we actually
do nonperturbative physics from the flat spacetime perspective.
Additionally, different backgrounds can be viewed as a resummation of collective gravitational fluctuations around an initial background.

On the other hand, the inclusion of background spacetimes of constant curvature is a very mild modification that can be treated exactly without tremendous efforts in computations. Therefore, it is an interesting laboratory to study perturbative implications of the same theory, but on different maximally symmetric backgrounds. For example, we can play with the value of the curvature radius of the background and easily we can check the claims about background-independence of nonlocal theories. Since AdS spacetimes gained in the last two decades a lot of attention, mainly due to the AdS/CFT conjecture, it is also highly desirable to have a gravitational version of nonlocal theories formulated on general AdS backgrounds. This could be viewed as a first step towards the investigation of the gauge-gravity duality in a class of weakly nonlocal gravitational theories consistent at quantum level.

The cosmological constant $a_\Lambda$ appearing in the tree-level action should be understood as a new coupling constant of gravitational character, and not like a special matter source.
Moreover, the cosmological constant term is an IR completion of the theory when we introduce all possible operators with a fixed number of derivatives. In the case of the cosmological constant we actually add a generally covariant term with no derivatives at all.
And then the following issue arises that for consistency we should quantize physical theory around on-shell background, i.e. such that solves exactly classical equations of motion. If we have the cosmological constant in the action, then the flat spacetime is not a solution anymore and we have to study the theory around Einstein spaces. The de Sitter and anti–de Sitter spacetimes serve as examples of such background spacetimes.

However, here we want to remark that it is also possible to pursue a different idea that the cosmological constant $\Lambda_{\rm cc}$ may not be present in the action.
%there for a spacetime background.
The background does not have to be on-shell with respect to the equations for the perturbations, and only at the end the physical theory for the on-shell Minkowski background without
$\Lambda_{\rm cc}$ should be considered.
If $\Lambda_{\rm cc}$ is in the action then the fluctuations must be analyzed around on-shell (A)dS spacetimes. On the other hand from the mathematical point of view it is consistent to have off-shell backgrounds on which one can have whatever theory describing the propagation and interactions of fluctuating modes.
We only want these fluctuations to be small (kind of a probe theory) to do not influence the background too much (backreaction is neglected). For example, we can study the quantum fluctuations
%a quantized setup for gravitational fluctuations
around the flat background in a theory that incorporates the cosmological constant term too. Indeed, we can always consider contributions to the propagator and vertices coming from the cosmological constant term even on a flat spacetime background.

%The stability of nonlocal theories around MSS (maximally symmetric spacetimes) is another interesting physical issue \cite{Khoury}. As remarked above not only the flat spacetime has the interest, but also all MSS should be considered and the existence of ghosts around these background is crucial for testing the good properties of the theory.
%{\color{red}However, the form of an action that gives rise to a nonlocal multiplicative modification of the Einstein-Hilbert propagator around MSS is known without ambiguities.} %There are many operators that should be taken into account, contrarily to the flat spacetime case where we have only 2 structures: $R \square^n R$ and ${\bf Ric} \square^n {\bf Ric}$. Here there are many more operators if we try to list them using standard curvature tensors (the rescue comes if we instead use the hatted curvature variables defined later in the paper and first proposed by Anselmi \cite{AnselmiQG}) Since generally not much was known about this class of theories on the (A)dS background it is worth to invest research in it. There is still a debate, whether we should care or not about the stability around  any MSS in this class of theories. Maybe it is enough to find a stable vacuum for the gravitational theory. However, it is still desired to investigate the stability around any MSS.

In this paper we show that the analysis of perturbative linear stability (equivalent to the analysis of the spectrum of linear perturbations around a given background) gives the same results as in the flat spacetime case and hence these gravitational vacuum configurations are perfectly stable. There is still a question, what is the vacuum here: gravitational vacuum or vacuum with a value of the cosmological constant, or no gravitational field at all (flat Minkowski spacetime). Besides this we think that having a one vacuum (which maybe even false) is fine for having a good candidate for quantum gravitational fundamental theory. Moreover, the quantization around MSS may have very important meaning in nonsupersymmetric theories (in unbroken supergravity the vacuum must be flat due to constraints coming from the supersymmetry algebra) and it must be considered seriously like the quantization around flat spacetime.

Last but not least,
%since observations in cosmology suggest that we are living in an exponentially expanding de-Sitter-like universe, then
%it is inevitable to have
we must look for a theory consistent on a
%with this fact and does not have any problem around the
MSS spacetime background because the cosmological
observations suggest that we are living in an exponentially expanding de-Sitter-like universe.
Despite that for all Earth- and solar system-based gravitational experiments we can safely neglect the effect of being in the dS phase, it is crucial %for the theoretical framework
that the theory, which has very good quantum properties around the Minkowski background, can be also formulated around any other MSS without major obstructions. Hereby, we show that such theory exists and is well defined, and it has the same analogous good quantum and UV properties as the theory previously studied on the flat spacetime background. The theory around any MSS possesses the following virtues: is generally-covariant, background-independent, perturbatively unitary \cite{Pius,Sen,Briscese:2018oyx,Chin}, and in the quantum domain can be easily selected to be superrenormalizable or UV finite. It is easily seen that the presence of one constant parameter (namely the cosmological constant) in this fundamental theory does not destroy, but rather only generalizes, the amazing structure already known around the flat background.
Indeed, on a MSS the theory is only slightly modified with respect to the theory on the flat background. %because the cosmological constant parameter $\Lambda_{\rm cc}$ enters in the game.
The MSS backgrounds are very well behaved: they are for example constant with respect to covariant derivatives and the commutators of derivatives can be traced back to some correction proportional to $\Lambda_{\rm cc}$. Moreover, the background curvature tensors can be completely written out using only the metric tensor and the parameter $\Lambda_{\rm cc}$. Therefore, we are going to include the cosmological constant in all the operators present in the action. In particular, the form factors could be selected to be functions of $\Lambda_{\rm cc}$ or the Ricci scalar. In the former case, as an additional advantage we can easily recover the flat spacetime results, by taking the limit
$\Lambda_{\rm cc}\to0$.

At the level of classical solutions the gravitational potential in the class of nonlocal theories is singularity-free and approaches a constant at $r=0$, regardless of the particular form factor appearing in the action %approximate solutions there are evidences endorse that we are dealing with a
%``{\em singularity-free gravitational}" for the case of physical matter
\cite{ModestoMoffatNico,Frolov1,Frolov2,Frolov3,Frolov4,Frolov5,Frolov6,Frolov7,Frolov8,BambiMalaModesto2, BambiMalaModesto,calcagnimodesto}. This was found in the context of approximate solutions. On the other hand regular bouncing solutions and the Starobinsky's cosmological solution have been shown to solve exactly the equations of motion of the nonlocal theory \cite{KoshelevStaro}. However, Ricci-flat spacetimes and the Friedmann-Robertson-Walker (FRW) spacetime in the presence of radiation are still exact solutions of the weakly nonlocal theory \cite{exactsol}. This issue has also to do with the question of localization of nonlocal theories as addressed in \cite{Cnl1}. Therefore, any form of nonlocality is not enough to smear out the singularities. However, at the present stage we cannot exclude that a special nonlocal theory could have only nonsingular solutions. Moreover, we have evidences that in this class of theories with infinitely many derivatives the black hole entanglement entropy is completely regularized and takes only finite values \cite{entanglement,myung}.

% different places of the action without exception whatsoever.
%
%Afterwards, we select out the entire functions so that the action quadratic in the graviton perturbation
%recasts in the linearization of the Einstein-Hilbert theory times two form factors for the traceless and trace parts of the graviton field respectively. This will be enough to get the gauge fixed propagator in (A)dS space.

%\section{Weakly nonlocal gravity without cosmological constant}

% \cor  summary of what is in each chapter of the paper\con

 In Sec. \ref{sectionWNL} we review the perturbative weakly nonlocal gravitational theory around the Minkowski space: the propagator, power counting, superrenormalizability, and finiteness at quantum level.
 In Sec. \ref{s3} we propose two classes of weakly nonlocal theories on (A)dS and we explicitly prove that the action at the second order in the graviton fluctuations has the same structure of the Einstein-Hilbert one. In Sec. \ref{s4} we show that the theory is finite at quantum level, while in Sec. \ref{GFR} we prove that
 for one out of the two classes of theories a perturbative field redefinition allows to map the nonlocal theory into the Einstein-Hilbert theory plus cosmological constant. In the last section we propose and study the most general weakly nonlocal field theory.

Most of the results obtained in this paper can be easily exported to Lee-Wick gravitational theories \cite{shapiro1, shapiro2, shapiro3,lm5,CLOP,HigherDG0} just replacing the nonlocal form factors with appropriate polynomials.

\section{nonlocal gravitational theories on Minkowski vacuum}\label{sectionWNL}

The most general $D$-dimensional theory weakly nonlocal (or quasilocal) and quadratic in curvature reads
\cite{Krasnikov,Kuzmin,Tombo,modesto,modestoLeslaw, Briscese:2013lna, M3, M4, Khoury, Mtheory,NLsugra},
\be
&&
\mathcal{L}_{\rm g} = -  2 \kappa_{D}^{-2} \, \sqrt{|g|}
\left[ R
+
 R \,
 \gamma_0(\Box)%_{\Lambda})
  R
 + {\bf Ric} \,
\gamma_2(\Box)%_{\Lambda})
 {\bf Ric}
+ {\bf Riem}  \,
\gamma_4(\Box)%_{\Lambda})
{\bf Riem}
+ {\cal V} \,
\right]  .
\label{gravityG}
\ee
The above Lagrangian density of the theory consists of a kinetic weakly nonlocal operator quadratic in curvature, three entire functions
$\gamma_0(\Box)$, $\gamma_2(\Box)$, $\gamma_4(\Box)$, and a set of local terms  ${\cal V}$ cubic or higher in curvature\footnote{{\em Definitions ---}
The metric tensor $g_{\mu \nu}$ has
signature $(- + \ldots +)$ and the curvature tensors are defined as follows:
$R^{\mu}{}_{\nu \!\rho \sigma} = - \partial_{\sigma} \Gamma^{\mu}{}_{\nu\!\rho} + \ldots $,
$R_{\mu \nu} = R^{\rho}{}_{\mu  \rho \nu}$,
$R = g^{\mu \nu} R_{\mu \nu}$. With symbol ${\cal R}$ we generally denote one of the above curvature tensors.}. The latter consists of operators with a properly chosen number of derivatives to not spoil the good
quantum properties of the theory.
% (in odd dimension we do not need to introduce any $\cal V$ to make the theory finite),
%The sum in (\ref{K0}) must include at least the minimal set of operators (with different tensorial structure), which we need to make the theory finite.
Moreover,
$\Box = g^{\mu\nu} \nabla_{\mu} \nabla_{\nu}$ is the covariant d'Alembertian (or box) operator, while the entire functions $\gamma_\ell(\Box)$
%$H(-\Box_{\Lambda})$
%will be shortly defined.
%The capital $\rm{N}$ is defined to be the following function of the spacetime dimension $D$: $2 \mathrm{N} + 4 = D$. \cor The REASONS for such modifications should be explained here \con
%The form factor $\gamma_i(\Box)$
%
are defined in terms of exponentials of entire functions $H_\ell(z)$ ($\ell=0,2$), namely
\be
&& \gamma_0(\Box) = - \frac{(D-2) ( e^{H_0(\Box)} -1 ) + D ( e^{H_2(\Box)} -1 )}{4 (D-1) \Box} + \gamma_4(\Box)
\label{gamma2} \, , \\
&& \gamma_2(\Box) = \frac{e^{H_2(\Box)} -1 }{\Box} - 4 \gamma_4(\Box) \, ,
\label{gamma0}
\ee
while $\gamma_4(\Box)$ stays arbitrary. It is only constrained by renormalizability to have the same asymptotic UV behavior as the other two form factors $\gamma_\ell(\Box)$ ($\ell=0,2$). The minimal choice
compatible with unitarity and superrenormalizability corresponds to  retaining only two out of three
form factors, i.e. we can choose $\gamma_4(\Box) =0$.

%As a matter of fact we can also add other operators quadratic in curvature and equivalent to the above operators
%up to interaction vertices. These operators correspond to a different ordering of covariant derivatives in introducing the form factors
%in-between the Riemann, Ricci, and scalar curvatures. As we said in the introduction here we consider theories on on-shell backgrounds, that's why on flat spacetime we do not add a cosmological constant term $\Lambda_{\rm cc}$ to the action.

Finally, the entire functions $V^{-1}_{\ell}(z) \equiv \exp (H_{\ell}(z))$ ($z \equiv - \Box_{\Lambda} \equiv - \Box/\Lambda^2$) (for $\ell=0,2$) introduced in (\ref{gamma2}) and (\ref{gamma0})
satisfy the following general conditions \cite{Tombo, modestoreview}:
\begin{enumerate}[(i)]
\item
$V^{-1}_{\ell}(z)$ is real and positive on the real axis and it has no zeros on the
whole complex plane $|z| < + \infty$. This requirement implies that there are no
gauge-invariant poles other than the transverse massless physical graviton pole;
\item
$|V^{-1}_{\ell}(z)|$ has the same asymptotic behavior along the real axis at $\pm \infty$;
\item
%{({\rm iii})}
There exist $\Theta>0$, $\Theta<\pi/2$ and positive integer $\gamma$, such that asymptotically
\be
&&
|V^{-1}_{\ell}(z)| \rightarrow | z |^{\gamma + \mathrm{N}+1},\,\, {\rm when }\,\, |z|\rightarrow + \infty
\quad {\rm with}
\quad
\gamma\geqslant \frac{D_{\rm even}}{2} \,
\quad  {\rm or}
\quad
 \gamma > \frac{D_{\rm odd}-1}{2} \, ,
\label{tombocond}
\ee
%(in order to avoid fractional powers of the d'Alembertian operator)
for the complex values of $z$ in the conical regions $C$ defined by:
$$C = \{ z \, | \,\, - \Theta < {\rm arg} z < + \Theta \, ,
\,\,  \pi - \Theta < {\rm arg} z < \pi + \Theta\}.$$
\end{enumerate}
The last condition is necessary to achieve the maximum convergence of the theory in
the UV regime.
The necessary asymptotic behavior is imposed not only on the real axis, but also on the conical regions, that surround it.
In an Euclidean spacetime, the condition (ii) is not strictly necessary if (iii) applies. In \eqref{tombocond} the capital $\rm{N}$ is defined to be the following function of the spacetime dimension $D$:
$2 \mathrm{N} + 4 = D_{\rm odd} +1$ in odd dimensions and $2 \mathrm{N} + 4 = D_{\rm even}$
in even dimensions. Moreover, by $\Lambda$ we denote the scale of nonlocality of the theory (not to be confused with the cosmological constant: $\Lambda_{\rm cc}$).

One example of such entire function due to Tomboulis \cite{Tombo} is:
\be
V^{-1}(z)= e^{\frac{1}{2} \left[ \Gamma \left(0, p(z)^2 \right)+\gamma_E  + \log \left( p(z)^2 \right) \right] } ,
\label{TomboFF}
\ee
where $\Gamma(0,x)$ is the incomplete Gamma function with its first argument put to zero, $p(z)$ is a polynomial of degree $\gamma+{\rm N} +1$ and $\gamma_E$ is the Euler-Mascheroni mathematical constant.
%The necessary asymptotic behavior is imposed not only on the real axis, but also on the conical regions, that surround it.
%In an Euclidean spacetime, the condition (ii) is not strictly necessary if (iii) applies.
%RElation between V_0 and V_2 and degrees of polynomials.
%For the situation with maximal convergence of the propagator of the theory
To achieve (super-)renormalizability %we must require
the degrees of the polynomials appearing in the definitions of $V^{-1}_0(z)$ and $V^{-1}_2(z)$ must be equal. % and this we will assume later in this paper.
In the rest of the paper we will denote %always assume the same dWe continue denoting
the common degree by $\gamma+{\rm N}+1$ ($N=0$ in $D=4$).

A few comments are in order here.
\begin{enumerate}[(i)]
	\item First, it is obvious that the Minkowski spacetime is indeed a solution of the background equations of motion (EOM) corresponding to the above action (\ref{gravityG}). Other terms than the Einstein-Hilbert one in the original Lagrangian are at least quadratic in curvature and as such vanish when evaluated for the Minkowski metric. Even though the EOM are not used in our present analysis the reader can find them in \cite{exactsol,Koshelev:2013lfm,meBiswasConroyMaz}. A cosmological constant term cannot be introduced here as it would lead to a constant nontrivial curvature at least.
	%%%
	\item Second, the action (\ref{gravityG}) is written exactly in the form as it is above because below we want to highlight the structure of the gravity propagator. Since the propagator can be read from a quadratic variation of the background action we worry about terms at most quadratic in curvatures. Higher curvature corrections vanish as upon the second variation as long as the background curvature itself is zero (as it is the case in Minkowski flat background).
	%%%%
	\item The requirement that the form factors $\gamma_\ell$ are entire functions deserves a little bit more explanation. The object of our consideration are \textit{weakly} nonlocal theories. This means that we have analytic function in the whole complex plane with in particular a smooth limit when momenta tend to zero. The way to think about this is to introduce a scale of gravity modification $\Lambda$ with the dimension of mass and carefully write everywhere $\Box/\Lambda^2$. As such the low energy limit is when the nonlocality scale goes to infinity. From here we find out that the form factors must be at least analytic in the origin. One may wonder why we need them to be entire functions i.e. analytic everywhere. It can be shown that the propagators of canonical variables in Arnowitt-Deser-Misner (ADM) formalism (which are observable quantities during inflation, for instance) feature a propagator with the form factor $\gamma_0$ in the denominator. As such, if the function $\gamma_0$ has some pole, it will become a new pole for the canonical variable and the quantum properties of the theory would be spoiled w.r.t. our expectations for the measurements. The mathematical details of this arguments can be found in a parallel study
\cite{staroNEW}. However, we make a statement that indeed the functions $\gamma_\ell$ must be entire.
%%%
	\item The advertised above form of the form factors $\gamma_\ell$ and the comment that only two out of three of these functions are essential is a consequence of the structure of the propagator and this is the matter of the succeeding analysis. It is however worth mentioning that the formulae (\ref{gamma2}), (\ref{gamma0}) do not guarantee that $\gamma_\ell$ are entire functions even though the functions $H_{0,2}$ are. This should be checked independently.
\end{enumerate}

We additionally remark here the reason to call
%that it is not really appropriate to call
the term $\mathcal{V}$ appearing in the action \eqref{gravityG} ``curvature potential''. First of all, we argue that for any gauge theory as well as for gravity the strict distinction between the kinetic term and the potential of interaction does not exist. This is due to gauge invariance that connects interactions also with standard terms responsible for the propagator. %propagation of fluctuations.
The nomenclature we have adopted here is that by kinetic terms we mean terms that do contribute to the propagator around flat spacetime.
%In the rest of the action we have terms, for which the propagator is independent of.
Around the flat spacetime typically the kinetic terms are operators up to quadratic in curvature, while in the ``curvature potential''  we put all the terms cubic and higher in the curvature. The counting above is insensitive to the number of covariant derivatives appearing in the term under consideration. This is the only meaningful difference between the two parts of the action. On MSS %we have terms, which are
operators cubic and higher in the curvature can contribute %contributing
to the propagator. % around this background.
However, we can suitably modify the potential to make it compatible with the above definition around the Minkowski flat spacetime.
Moreover, the locality of $\mathcal{V}$ is not a must, while the weak nonlocality is not required by the unitarity.
%It is crucial to understand that by investigating this kind of ``curvature potential" we can almost never derive clues about the stability of the theory or of it particular configurations as classical solutions. We elucidate about this below.
%

Finally, since in the gravitational case the notion of local energy density of the gravitational field is not well defined (strictly this is not a gauge-invariant observable with respect to the diffeomorphism group) we can not sensibly speak about the potential energy for the gravitational Lagrangian case. We want to emphasize that even in the case of finite QED (studied in \cite{piva}) the role of the potential $\cal V$ is different from the standard role ascribed to it in classical mechanics or in other field theory models, so it is a little inappropriate to call it that.

\subsection{Propagator and unitarity around the Minkowski spacetime}
Splitting the spacetime metric into the flat Minkowski background and the dimensionful fluctuation $h_{\mu \nu}$
defined by $g_{\mu \nu} =  \eta_{\mu \nu} + \kappa_D \, h_{\mu \nu}$ (here and above $\kappa_D$ is proportional to the square root of the gravitational Newton constant),
we can expand the action (\ref{gravityG}) to the second order in $h_{\mu \nu}$.
The result of this expansion together with the usual harmonic gauge-fixing term reads \cite{HigherDG}
\be
\mathcal{L}_{\rm quad} + \mathcal{L}_{\rm GF} = \frac{1}{2} h^{\mu \nu} \mathcal{O}_{\mu \nu, \rho \sigma} \, h^{\rho \sigma},
\label{O}
\ee
where the operator %inverse of the propagator
$\mathcal{O}$ is made out of two terms, one coming from the quadratization of (\ref{gravityG})
and the other from the following gauge-fixing term,
$\mathcal{L}_{\rm GF}  = \xi^{-1}  \partial^{\nu}h_{\mu \nu} \omega(-\Box_{\Lambda}) \partial_{\rho}h^{\rho \mu}$, where
% \label{GF2}
$\omega( - \Box_{\Lambda})$ is a weight functional \cite{Stelle, Shapirobook}.
The d'Alembertian operator in $\mathcal{L}_{\rm quad}$ and the gauge-fixing term must be conceived on %relative
%to
the flat spacetime.
Inverting the operator $\mathcal{O}$ \cite{HigherDG} and making use of the
form factors (\ref{gamma2}) and (\ref{gamma0}), we find the %following
two-point function in the harmonic gauge ($\partial^{\mu} h_{\mu \nu} = 0$),
\be
\mathcal{O}^{-1} = \frac{\xi (2P^{(1)} + \bar{P}^{(0)} ) }{2 k^2 \, \omega( k^2/\Lambda^2)}
%&& \hspace{-0.4cm}
+
\frac{P^{(2)}}{k^2   e^{H_2(k^2/\Lambda^2)} }
\label{propagator}
%&& \hspace{-0.5cm}
- \frac{P^{(0)}}{  \left( D-2 \right)k^2   e^{H_0(k^2/\Lambda^2)}} \,  .
\ee
%
 %\\
%&& \hspace{-0.4cm}
%+ \frac{\zeta}{2 k^2 \, \omega_1( k^2/\Lambda^2) (2P^{(1)} + \bar{P}^{(0)}. }
%\nonumber \\
We omitted the tensorial
indices for the propagator $\mathcal{O}^{-1}$ and the projectors $\{ P^{(0)},P^{(2)},P^{(1)},\bar{P}^{(0)}\}$
%The projectors are
defined in  %We have also introduced the following projectors
\cite{HigherDG, VN}\label{proje2}\footnote{The standard projectors are defined by:
%%%%%
\be
 && \hspace{-1.0cm}
 P^{(2)}_{\mu \nu, \rho \sigma}(k) = \frac{1}{2}( \theta_{\mu \rho} \theta_{\nu \sigma} +
 \theta_{\mu \sigma} \theta_{\nu \rho} ) -  \frac{1}{D-1} \theta_{\mu \nu} \theta_{\rho \sigma}
 \, ,
 \,\,\,\,  \,
 %\nonumber \\
% && % \hspace{-0.2cm}
P^{(1)}_{\mu \nu, \rho \sigma}(k) = \frac{1}{2} \left( \theta_{\mu \rho} \omega_{\nu \sigma} +
 \theta_{\mu \sigma} \omega_{\nu \rho}  +
 \theta_{\nu \rho} \omega_{\mu \sigma}  +
  \theta_{\nu \sigma} \omega_{\mu \rho}  \right)  \,  , \nonumber   \\
   &&
  \hspace{-1.0cm}
P^{(0)} _{\mu\nu, \rho\sigma} (k) =  \frac{1}{D-1}\theta_{\mu \nu} \theta_{\rho \sigma}  \, ,  \,\,\,\, \,
\bar{P}^{(0)} _{\mu\nu, \rho\sigma} (k) =  \omega_{\mu \nu} \omega_{\rho \sigma} \,  , % \nonumber  \\
\,\,\,\,
%&& %\hspace{-1.4cm}
\theta_{\mu \nu} = \eta_{\mu \nu} - \frac{k_{\mu } k_{\nu }}{k^2} \, , %\nonumber \\
\,\,\,\,
%&& \hspace{-1.4cm}
\omega_{\mu \nu } = \frac{k_{\mu} k_{\nu}}{k^2} \, .
  \label{proje3}
\ee
We have also replaced $-\Box$ by $k^2$ in the quadratized action.
%where
% $\theta_{\mu \nu} = \eta_{\mu \nu} - k_{\mu } k_{\nu }/k^2$ and $\omega_{\mu \nu } = k_{\mu} k_{\nu}/k^2$.
}

The propagator (\ref{propagator}) is the most general one compatible with unitarity. It propagates no other degree of freedom (d.o.f.) besides the standard massless transverse spin-2 graviton. This follows from the fact that exponents of entire functions are special entire functions with no zeros. So we technically avoid new poles which means we avoid new physical d.o.f.
Returning to the comment in the previous subsection we see that the structure of the propagator advocates the form of form factors $\gamma_\ell$ as the absence of new d.o.f. was exactly the requirement behind formulae (\ref{gamma2}), (\ref{gamma0}). We also note that in order to have a well behaved propagator we need to get a correct form of only two factors corresponding to spin-0 and spin-2 parts. This explains why one function out of three $\gamma_\ell$ can be put to zero from the point of view of unitarity.
Further, the unitarity is manifest, because the optical theorem at  tree-level is trivially satisfied, namely
 \be
 2 \, {\rm Im} \left\{  T(k)^{\mu\nu} \mathcal{O}^{-1}_{\mu\nu, \rho \sigma} T(k)^{\rho \sigma} \right\} = 2  \pi \, {\rm Res} \left\{  T(k)^{\mu\nu} \mathcal{O}^{-1}_{\mu\nu, \rho \sigma} T(k)^{\rho \sigma} \right\} \big|_{k^2 = 0}> 0,
 \ee
 where $T^{\mu\nu}(k)$ is the  Fourier transform of the conserved energy tensor  of a matter source.
%
%\subsection{{\bf nonlocal Propagators for the Graviton and Ghosts}}
 %if any \dots
%
%

 So far we have proved that the theory is unitary at perturbative level around the Minkowski spacetime. However, we will probably be able to prove the nonperturbative unitarity of the theory around the Minkowski spacetime in a soon future. On the other hand, unitarity around general backgrounds, is a very difficult task. Indeed, unitarity is only well defined in Minkowski spacetime mainly because it is not clear how to define the concept of particle in curved spacetime and even how unambiguosly define the asymptotic states for the scattering $S$ matrix. We remind that we can sensibly speak about unitarity of the $S$ matrix only in theories where we have a well defined $S$ matrix. Regarding issues related to unitarity, at most what can be proved is the absence of ghosts on any background for some special theories (for example Einstein gravity is ghost-free around any background). Nevertheless, in a recent paper we proved that we can have up to eight degrees of freedom in nonlocal gravity on a general background, and we do not know if some of them are ghostlike. In this paper we are going to prove that there are no ghosts in AdS and dS spacetimes (Sec. V), while in two other papers \cite{CalcagniModesto,CalcagniModestoMyung} we have proved the absence of ghosts (actually linear stability) around Ricci-flat spacetimes. However, the linear stability of a general background is really a difficult task and beyond the scope of this paper.

\subsection{%Propagator, tree-level unitarity and
Power counting in a nutshell } %, unitarity, and quantum divergences} %and finite theories}
\label{gravitonpropagator}

%Splitting the spacetime metric into the Minkowski background plus a fluctuation,
%after gauge fixing we can invert the quadratic operator to finally get the wo point function  \cite{HigherDG},
%The propagator in the Fourier space, up to gauge dependent components, reads
%\be
%&& \hspace{-1cm}   \mathcal{O}^{-1} \!=\!
 %\frac{V(  k^2/\Lambda^2 )  } {k^2} \left( P^{(2)} - \frac{P^{(0)}}{D-2 }  \right)
  %\, \times \,({\rm TENSOR - STRUCTURE})  .
%+  \frac{\xi (2P^{(1)} + \bar{P}^{(0)} ) }{2 k^2 \, \omega( k^2/\Lambda^2)} .
% \label{propagator}
%\ee
%The tensorial structure for gravity is
%${\rm TS} = \left( P^{(2)} - \frac{P^{(0)}}{D-2 }  \right)$.
%The indices for the operator $\mathcal{O}^{-1}$ and the projectors \cite{HigherDG, VN} $\{ P^{(0)},P^{(2)}%,P^{(1)},\bar{P}^{(0)
%\}$ have been omitted. %\footnote{
 %
%The tensorial structure in (\ref{propagator}) is the same of Einstein's gravity, but the multiplicative
 %form factor $V(-\Box_{\Lambda})$ makes the theory strongly UV convergent without the need to modify the spectrum or introducing ghost instabilities.

We now review \cite{Kuzmin,Tombo, TomboStudent, modesto, modestoLeslaw, universality} power counting analysis of the quantum divergences. We remark that the divergences do not depend on the choice of the background spacetime metric; therefore, the results in this subsection apply equally well to the case of theories studied around general MSS backgrounds.
In the high energy regime,
the above propagator (\ref{propagator}) in momentum space
schematically scales as
\be
\mathcal{O}^{-1}(k) \sim \frac{1}{k^{2 \gamma +D} } \quad % \,\,\,\,\,\,
\mbox{in the UV} \, .
\label{OV}
\ee
The vertices can be collected in different sets, that may or may not involve  the entire functions $\exp H_\ell(z)$.
However, to find a bound on the quantum divergences it is sufficient to concentrate on
the leading operators in the UV regime.
These operators scale as the inverse of the propagator giving the following
upper bounds on the superficial degree of divergence of any graph $G$ \cite{Kuzmin,Tombo,modesto,modestoLeslaw},
\be
\omega(G)=DL+(V-I)(2 \gamma + D),
\ee
in a spacetime of even or odd dimension respectively. We simplify the above relation further to
\be
&&
\omega(G) =  D - 2 \gamma  (L - 1)    \, . \,\,\,\,
\label{even}
\ee
In (\ref{even}),  we used the topological relation between the numbers of vertices $V$, internal lines $I$ and the number of loops $L$: $I = V + L -1$.
Thus, if $\gamma > D/2$, in the theory only 1-loop divergences survive.
Therefore, the theory is superrenormalizable \cite{Kuzmin,Tombo,modesto,modestoLeslaw, universality}
and only a finite number of operators of mass dimension up to $M^D$ has to be
included in the action in even dimension for the purpose of renormalization.

Notice that the power counting analysis can be done in Minkowski spacetime %not only because it is simpler, but
because any smooth spacetime is locally flat and the divergences are related to the UV coincidence limit in the correlation functions. Therefore, we can %evaluate the beta function
expand around whatever background, and we will always end up with the same divergent contributions to the quantum effective action; namely we will always get the same beta functions.

\subsection{The theory in Weyl basis}
We can equally consider a different action, which will be written by re-shuffling quadratic in curvature terms in (\ref{gravityG}). The following action
is equivalent to \eqref{gravityG} for
everything about unitarity [the propagator is given again by (\ref{propagator})] and superrenormalizability
or UV finiteness and its Lagrangian density reads
\be
&& \mathcal{L}_{\rm C} = -  2 \kappa_D^{-2} \sqrt{|g|}\Big[ { R} +
 {\bf C} %\mu \nu \rho \sigma}
  \gamma_{\rm C} (\Box) {\bf C}
  %C^{\mu \nu \rho \sigma}
 + { R}  \gamma_{\rm S}(\Box) {R }
 + {\bf Riem} \,  \gamma_{\rm R}(\Box) {\bf Riem} + {\cal V}
 \Big]  \, , %\quad  {\rm with}
 \label{TWeyl}
 \\
&&
\gamma_{\rm C} = - \frac{D-2}{4}  \gamma_2 \, , \quad % \nonumber \\&&
\gamma_{\rm S} = \gamma_0  +  \frac{1}{2(D-1)}  \gamma_2 \, , \quad
%\nonumber \\&&
\gamma_{\rm R} = \gamma_4 + \frac{D-2}{4}  \gamma_2 \, ,
\label{TWeylgamma}
\ee
where ${\rm \bf C}$ is the Weyl tensor and all the form factors $\gamma_{\ell}$ are defined in (\ref{gamma2}) and (\ref{gamma0}).
%The potential term $\mathcal V$ however is not necessarily equivalent to the one in action (\ref{gravityG}). We  may wonder considering \cor different structure of vertices ? here ? \con in order to select a gravity theory obeying certain renormalizability properties.
%
%After plugging their definitions to the above form factors in (\ref{TWeylgamma})
%we get
%$\gamma_{\rm C}$, $\gamma_{\rm S}$, and $\gamma_{\rm R}$
%\be
%&& \gamma_{\rm C} = \frac{(D-2) \left(-e^{H_2}+4 \gamma _4 \Box+1\right)}{4 \Box}
% \, , \\
%&& \gamma_{\rm S} =\frac{(2-D) \left(e^{H_0}+e^{H_2}-2\right)+4 \gamma _4 (D-3) \Box }{4 (D-1) \Box}
% \, , \quad
%  \gamma_{\rm R} = % \frac{(D-2) \left(e^{H_2}-1\right)-4 \gamma _4 (D-3) z}{4 z} \quad
%\frac{(D-2) \left(e^{H_2}-1\right)-4 \gamma _4 (D-3) \Box}{4 \Box} \, .
%\ee
%%

To start with, we recall that only two form factors are needed to have appropriate propagator. We thus put
$\gamma_{\rm R}=0$ and the theory (\ref{TWeyl}) reduces to:
\be
&& \mathcal{L}_{\rm C} = -  2 \kappa_D^{-2} \sqrt{|g|}\Big[ {R} +
 {\bf C} %\mu \nu \rho \sigma}
  \gamma_{\rm C} (\Box) {\bf C}
  + {R}  \gamma_{\rm S}(\Box) {R }
  %C^{\mu \nu \rho \sigma}
  + {\cal V}({\bf C})
 \Big]  \,  ,
 \label{TWeyl2}
 \\
&&
\gamma_{\rm C} =   \frac{D-2}{4(D-3)}  \frac{e^{H_2} -1}{\Box} \, , \quad
\gamma_{\rm S} = - \frac{D-2}{4(D-1)}  \frac{e^{H_0} -1}{\Box} \, .
\label{formMink}
\ee
%We can consider a different action in the Weyl basis, which is equivalent to the previous one for everything about unitarity and superrenormalizability. The action reads,
%\be
%&& \mathcal{L}_{\rm C} = -  2 \kappa_D^{-2} \sqrt{|g|}\Big[ R +
 %\frac{1}{2}C_{\mu \nu \rho \sigma} h_c(-\Box_{\Lambda}) C^{\mu \nu \rho \sigma} - R  h_R(-\Box_{\Lambda}) R \Big]  \,  ,
 %\nonumber
 %\\
%&&
%h_c = \frac{e^{H(-\Box_{\Lambda}) } -1}{\Box} \,\, , \,\,\,\, h_R =  \frac{1}{6} h_c  \, .
%\ee
%The propagator is given in (\ref{propagator}) with the pole in $k^2 = 0$.
%This theory will also likely  turn out to be finite at quantum level, if a specifically chosen set of local terms in $\cal V$ is included in the action.  We can here always build it up with only Weyl tensor, namely ${\cal V}={\cal V}({\bf C})$.
%
In $D=4$ it is enough to include $\cal V$
made out of two Weyl killers to end up with a completely finite quantum gravitational theory at any perturbative order in the loop expansion. For example we can choose the following two operators,
\be
{\cal V}({\bf C}) = s^{(1)}_w \, C_{\mu\nu\rho\sigma} C^{\mu\nu\rho\sigma} \Box ^{\gamma -2} C_{\alpha \beta \gamma \delta}
C^{\alpha \beta \gamma \delta} +
s^{(2)}_w \, C_{\mu\nu\rho\sigma} C^{\alpha \beta \gamma \delta} \Box ^{\gamma -2} C_{\alpha \beta \gamma \delta}
C^{\mu\nu\rho\sigma} \, .
\label{killersWtext}
\ee
The Gauss-Bonnet (GB) operator does not contribute to the divergent part of the quantum effective action in $D=4$ when the manifold has a trivial topology, namely the spacetime is topologically equivalent to the Minkowski one or the Euclidean space
(see for example \cite{Kuzmin}).
However, in the rest of the paper we will deal with the (A)dS space and we will have to take care of the
divergence proportional to the GB too.

The beta functions for the two couplings in front of terms quadratic in curvature can be only linear in the front coefficients $s^{(1)}_w$ and $s^{(2)}_w$ \cite{modestoLeslaw}, then we can always find a solution
to the equations $\beta_{ R^2} =0$ and $\beta_{\bf Ric^2} =0$ regardless of the energy scale and the loop order. The integral of the Gauss-Bonet operator is in this section identically zero because we assume the space to be topologically equivalent to the Minkowski spacetime. Later we will be forced to give up this hypothesis in (A)dS.
%{\color{red}See later for more details.}

As pointed out in the introduction the weak nonlocality is not sufficient to solve the singularity issue that plagues the Einstein-Hilbert gravitational theory. In particular, for the theory in the Weyl basis presented in this section the FRW metrics for conformal matter ($T_{\rm matter} \equiv 0$) solve exactly the nonlocal EOM \cite{exactsol}. This means that the Big-Bang singularity shows up in an exact solution of our nonlocal quantum gravity. However, if the gravitational sector also enjoys conformal invariance, then any FRW singular spacetime is conformally equivalent to the flat spacetime by a conformal rescaling and the singularity turns out to be unphysical \cite{ModestoLeslawConformal}. Notice, that the presence of singularities in particular nonlocal theories does not rule out that it may exist a nonlocal theory, which is singularity-free. However, the naive nonlocality by itself is not enough \cite{exactsol,ModestoLeslawConformal}. The crucial ingredient here is the conformal symmetry, which allows for rescalings like described above. Moreover scale invariance helps with singularity of geodesics \cite{ModestoLeslawConformal,bambi1} and also with some issues of black hole physics \cite{bambi2,Myung:2017zyb}.

%% \subsection{The theory in Einstein basis}
%Another basis, in which we express the theory is the one introduced in \cite{Tombo, modesto,Barvinsky}.
%Letting $\gamma_4 =0$ in (\ref{gravityG}) we find
%\be
%&& \mathcal{L}_{\rm E} = -  2 \kappa_D^{-2} \sqrt{|g|}\Big[ { R}+
% {\bf G} %\mu \nu \rho \sigma}
%\,   \gamma_{\rm G} (\Box) {\bf Ric}
%  %C^{\mu \nu \rho \sigma}
% + { R}  \gamma_{\rm S^{\prime}}(\Box) { R }
%  + {\cal V}
% \Big]  \quad {\rm with}
% \nonumber
% \\
%&&
%\gamma_{\rm G} =  \gamma_2 =  \frac{e^{H_2} -1 }{\Box} \, , \quad
%\gamma_{\rm S^{\prime}} = \frac{1}{2} \gamma_2 + \gamma_0 = \frac{D-2}{4(D-1)} \frac{e^{H_2} - e^{H_0}}{\Box} \, .
%\label{EinsteinBasis}
%\ee
%We can analyze a minimal choice $H_0=H_2$ when the theory reduces to
%\be
%&& \mathcal{L}_{\rm E} = -  2 \kappa_D^{-2} \sqrt{|g|}\Big[ { R} +
% {\bf G} %\mu \nu \rho \sigma}
%\,   \gamma_{\rm G} (\Box) {\bf Ric}
%  %C^{\mu \nu \rho \sigma}
%  + {\cal V}
% \Big]  \quad {\rm with} \quad
% \quad \gamma_{\rm G} = \frac{e^{H_2} -1 }{\Box} \, .
%\label{EinsteinBasisS}
%\ee

\section{nonlocal gravity in (A)dS vacuum} \label{s3}
A generalization to a constant curvature background is rather straightforward.
We here provide the expansion of the action to the second order in the gravitational fluctuations around an (A)dS spacetime and we will infer about the stability properties of the theory around any maximally symmetric vacuum.
We retrace the path followed for the case of the Minkowski vacuum to mimic as much as possible
the Einstein-Hilbert theory. Therefore, we will end up with a quadratic operator that reproduces the one from Einstein's gravity on the same background up to, at most, two multiplicative form factors that do not change the structure of the classical two-point function \cite{Allen}.  Technically, we will use the previous computations published in \cite{meBiswasMaz,KoshelevStaro,meBiswasMazLONG} (see also Appendix). For definiteness we will first concentrate on the situation on on-shell MSS backgrounds. For them we can use that $a_\Lambda=\Lambda_{\rm cc}$.

For reasons that will be clear later in the paper, we here study two classes of theories that we identify as theories in the ``Weyl basis" and theories in the ``Ricci basis".

\subsection{A class of theories in the Weyl basis}
To see how things work we stick to $D=4$, make use of the Weyl basis, and consider the case $\gamma_R=0$,
\be
&&\boxed{ \mathcal{L}_{\rm CR} = -  2 \kappa_4^{-2} \sqrt{|g|}\Big[ { R} - 2 \Lambda_{\rm cc} +
 {\bf C} \, %\mu \nu \rho \sigma}
  \gamma_{\rm C} (\Box) {\bf C}
  + { R}  \gamma_{\rm S}(\Box) { R }
  %C^{\mu \nu \rho \sigma}
  + {\cal V}({\bf C})
 \Big]
 }
 \quad {\rm with}
 \label{TWeyl3}
 \\
&&
\boxed{\gamma_{\rm C}(\Box) =   \frac{1}{2}   \left( e^{H_2 \left( \Box -  \frac{2}{3} R \right) } -1 \right) \frac{1}{  \Box - \frac{2}{3}  R }  \, , \qquad
\gamma_{\rm S}(\Box) = - \frac{1}{6}   \frac{1}{  \Box+ \frac{R}{3}  }
\left( e^{H_0 \left( \Box, R \right) %+ \frac{R}{3} \right)
}  - 1    \right)
}
\, ,
\label{formADS}
\ee
where the translations of the  covariant box operators (by the amount proportional to $R$)
in comparison to the form factors given in (\ref{formMink}) will be shortly clear.
Notice that the form factors (\ref{formADS}) turn in (\ref{formMink}) when the formal limit
$R \rightarrow 0$ in (\ref{formADS}) is taken. Moreover, the ordering of the operators could be relevant in (\ref{formADS}) for some choices of the asymptotic polynomials. Indeed, the arguments of the
entire functions $H_0$ and/or $H_2$ can in general differ from the denominators in
(\ref{formADS}) so that they do not commute due to the Ricci scalar curvatures appearing with different numerical coefficients\footnote{Another slightly different choice of the form factors with respect to
(\ref{formADS}) can make irrelevant the ordering, namely
\be
\gamma_{\rm C}(\Box) =   \frac{1}{2} \frac{  e^{H_2 \left( \Box -  \frac{8}{3} \Lambda_{\rm cc} \right) } -1  }{  \Box - \frac{8}{3}  \Lambda_{\rm cc}  }  \, , \quad
\gamma_{\rm S}(\Box) = - \frac{1}{6}  \frac{ e^{H_0 \left( \Box+ \frac{4}{3} \Lambda_{\rm cc} \right)}  - 1     }{  \Box + \frac{4}{3} \Lambda_{\rm cc}  }
\, ,
\label{formADS2}
\ee
where we replaced $R$ with the cosmological constant that now appears not only in the local
Einstein-Hilbert sector of the theory, but also explicitly in the form factors. Notice that this is an off-shell replacement, it is just a different definition of the theory. %\cor %However
Here the amount of the shift in terms of the cosmological constant had been fixed in order
to have stability around the (A)dS spacetime. Moreover, the form factors (\ref{formADS2})
can be easily expressed as $\sum_{r=0}^{\infty} a_r \Box^r$ for a proper choice of the coefficients $a_r$ because $\Lambda_{\rm cc}$ is a constant and then the form factor is commutative contrary to the one in the main text, namely (\ref{formADS}). Therefore, it is easy to implement the power counting analysis developed in \cite{Kuzmin,Tombo,TomboStudent}. The vertices for the theory with form factors (\ref{formADS2}) will contain the incremental ratios defined in \cite{TomboStudent} for the same form factors (\ref{formADS2}) with $\Box$ replaced with $\Box_M$ (the box operator on Minkowski space) (see also the discussion in the last part of this subsection).
% from solution of exact EOM.
%\con
}.

In taking the quadratic part of the action (\ref{TWeyl3}) in the graviton fluctuation $h_{\mu\nu}$ we use the following decomposition of the graviton field,
\be
h_{\mu\nu} = h_{\mu\nu}^{\bot} + \nabla_{(\mu} A^\bot_{\nu)} +
\left( \nabla_\mu \nabla_\nu - \frac{1}{4} g_{\mu\nu} \Box \right) B + \frac{1}{4} g_{\mu\nu} h ,
\ee
where the spin-two fluctuation $h_{\mu\nu}^{\bot}$ contains $5$ degrees of freedom because
it satisfies $\nabla^\mu  h_{\mu\nu}^{\bot} = g^{\mu\nu} h_{\mu\nu}^{\bot} =0$. The transverse vector
$A^\bot_{\nu}$,  satisfying $\nabla^\mu A^\bot_{\mu}=0$, is  accounting for three degrees of freedom. Finally, $B$ and $h$ are two real scalars. However, $A^\bot_{\mu}$ automatically drops out of the second variation of the action and out of the two scalars only the following combination
$\phi = \Box B - h$ appears there.

We end up with the following second order variation of the action \cite{meBiswasMaz,KoshelevStaro,meBiswasMazLONG},
\be
&& S^{(2) {\rm CR} }_{\rm (A)dS} = \frac{1}{2} \int d^4 x \sqrt{| \bar{g}|} \left\{ \tilde{h}^{\bot \mu\nu}
\left( \Box - \frac{\bar{R}}{6} \right) \left[ 1 + 2 \gamma_{\rm S}(0) \bar{R}
+ 2 \left( \Box - \frac{\bar{R}}{3} \right) \gamma_{\rm C}\left( \Box + \frac{\bar{R}}{3} \right)
\right] \tilde{h}^{\bot}_{ \mu\nu} \right. \nonumber \\
&& \left.
\hspace{5.2cm}
- \tilde{\phi} \left( \Box + \frac{\bar{R}}{3} \right) \left[ 1 + 2 \gamma_{\rm S}(0) \bar{R}
-  6 \left( \Box + \frac{\bar{R}}{3} \right) \gamma_{\rm S}( \Box )
\right] \tilde{\phi} \right\} \, ,
\label{variationADS}
\ee
where we introduced the canonically normalized fields
$\tilde{h}^{\bot}_{ \mu\nu} = M_{\rm P} \,  {h}^{\bot}_{ \mu\nu}/2$,
$\tilde{\phi} = \sqrt{3/32} \, M_{\rm P} \, \phi$, and $M^2_{\rm P} = 4 \kappa_4^{-2}$. In this part of the section the bar operators $\bar{\mathcal{O}}$ denote any background quantity.
Moreover, $\gamma_{\rm S}(0)$ can be read out of the following general expansion,
%we mean that there is no constant term in $\gamma_{\rm S}$, namely
\be
\gamma_{\rm S} (\Box) = \sum_{i=0}^{+\infty} c_{{\rm S}, i} \left[ \left(\Box + X \right)^{n_2} \right]^i \left( \Box^{n_1} \right)^i \, , \quad n_1,n_2 \in \mathbb{N} \, , %\quad c_{{\rm S},0} =0 \, ,
\label{traslato}
\ee
where $X$ is an operator proportional to the background Ricci scalar $R$. The chosen order in (\ref{traslato}) is consistent with the polynomial given below in  (\ref{ps}). %In short we require $\gamma_{\rm S} (\Box=0)=0$.
%
%\be
%\gamma_{\rm S} = \sum_{r=0}
%\ee
%
The detailed expressions for %such background operators in terms of the cosmological constant and
the most general second order variations of various actions on MSS are collected in Appendix.
We can assume $\gamma_{\rm S}(0) = 0$ (i.e. $c_{{\rm S},0} =0$ in (\ref{formADS}), (\ref{traslato})) because this is consistent with the requirements for the special entire
function $H(z)$ \cite{Tombo}. Therefore, replacing the form factors (\ref{formADS}) in the variation
(\ref{variationADS}) we end up with the following result,
\be
 S^{(2)  {\rm CR}}_{\rm (A)dS} = \frac{1}{2} \int d^4 x \sqrt{|\bar{g}|} \left\{ \tilde{h}^{\bot \mu\nu}
\left( \Box - \frac{\bar{R}}{6} \right) \, e^{H_2 \left( \Box - \frac{\bar{R}}{3} \right)}
\, \tilde{h}^{\bot}_{ \mu\nu}
- \tilde{\phi} \left( \Box + \frac{\bar{R}}{3} \right) \, e^{H_0 \left( \Box, \bar{R}
%H_0 \left( \Box+ \frac{\bar{R} }{3}
\right) } \, \tilde{\phi} \right\}.
\label{variationADS2}
\ee
The condition $\gamma_{\rm S}(0) = 0$ and the locality of counterterms force us to
select the following entire function (we here consider the $\gamma=3$ case),
\be
 && H_0 \left( \Box , R \right)
  %+ \frac{{R}}{3} \right)
 = \frac{1}{2} \left\{ \gamma_E
+ \Gamma \left( 0 , \left[  p_{\rm S} \left( \Box, R \right) %+ \frac{{R}}{3} \right)
\right]^2 \right)
+ \log \left[  p_{\rm S} \left( \Box, R \right) %+ \frac{{R}}{3} \right)
\right]^2 \right\} \, , %\quad %
\\
&&
p_{\rm S}\left( \Box, R \right)
%+ \frac{{R}}{3} \right)
= \frac{1}{\Lambda^8}  \left( \Box+ \frac{{R}}{3} \right)^2 \Box^2 \, .
%\qquad
%\color{red}{ p_{\rm S} \left( \Box+ \frac{{R}}{3} \right) = \frac{1}{\Lambda^8} \left( \Box+ \frac{{R}}{3} \right)^4 }.
\label{ps}
\ee
%Here by $\gamma_{\rm S}(0)$ we mean $\gamma_{\rm S}( \Box+ R/3=0)$ because the argument is $\Box+ R/3$.
For the form factor $\gamma_{\rm C}$ we can take the following entire function $H_2$,
\be
&& H_2 \left( \Box -  \frac{2}{3}  R \right) % + \frac{{R}}{3} \right)
 = \frac{1}{2} \left\{ \gamma_E
+ \Gamma \left( 0 , \left[  p_{\rm C} \left( \Box -  \frac{2}{3}
 R \right) %+ \frac{{R}}{3} \right)
\right]^2 \right)
+ \log \left[  p_{\rm C} \left( \Box %,R \right)
-  \frac{2}{3}  R \right) %+ \frac{{R}}{3} \right)
\right]^2 \right\} \, ,
\nonumber \\
&&
 p_{\rm C} \left( \Box -  \frac{2}{3}
 R \right) = \frac{1}{\Lambda^8} \left( \Box -  \frac{2}{3} R \right)^4 .
\label{pc}
\ee
%%%%%% Attenzione
We notice here that the choice of polynomials is fixed only by the UV behavior of the propagator and as such we have a lot of freedom in choosing them as long as basic principles are obeyed.

For instance we can have ``commutative" form factors (\ref{formADS2}) such that
\be
&&
p_{\rm C}\left( \Box, \Lambda_{\rm cc} % - \frac{8}{3} \Lambda_{\rm cc}
 \right) = \frac{1}{\Lambda^8} \Box^2 \left( \Box-  \frac{8}{3}  \Lambda_{\rm cc} \right)^2  \, , \\
&&
p_{\rm S}\left( \Box, \Lambda_{\rm cc}
%\left( \Box+ \frac{{4}}{3} \Lambda_{\rm cc}
 \right) = \frac{1}{\Lambda^8}  \Box^2 \left( \Box+ \frac{{4}}{3} \Lambda_{\rm cc} \right)^2 \, .
\label{ps2}
\ee
Notice that the above polynomials are zero for $\Box=0$, which is crucial to secure
%because $\Lambda_{\rm cc}$ is a constant. % and by
$\gamma_{\rm S}(0)=0$. % we mean $\gamma_{\rm S}(\Box =0)$.
Indeed, (\ref{ps2}) is a polynomial in $\Box$ and if we do not multiply by $\Box^2$ we get a constant dimensionless contribution proportional to $(\Lambda_{\rm cc}/\Lambda^2)^2 \propto c_{{\rm S},0}\neq 0$.
%but we are not forced to require such property because the $\Box$ operators in the denominators of $\gamma_{\rm C}$ and $\gamma_{\rm S}$ in
%(\ref{formADS2}) are translated of $- 8 \Lambda_{\rm cc}/3$ and $-4 \Lambda_{\rm cc}/3$ respectively.
%Notice that the un-translated $\Box$ operator is now crucial to avoid the constant term in
%$\gamma_{\rm  S}$.
%For the form factor $\gamma_C$ we can take the following polynomial

The following special choice of the polynomial $p_{\gamma +1}$ ($\gamma+1 =8$) makes also consistent the identification of $H_2$ with $H_0$, %. The polynomials for $\gamma+1 =8$ are given by
 \be
  {p}_8\left( \Box, R %+ \frac{{R}}{3}
  \right)  = \frac{1}{\Lambda^{16}}  \left( \Box+ \frac{{R}}{3} \right)^2  \Box^2 \left( \Box - \frac{2}{3} {R} \right)^4 \, .
 % \qquad
 %  \color{red}{{p}_8\left( \Box, R %+ \frac{{R}}{3}
%   \right)  = \frac{1}{\Lambda^{16}} \left[ \left( \Box+ \frac{{R}}{3} \right) \left( \Box - \frac{2}{3} {R} \right) \right]^4} .
 \label{poly8}
 \ee
 It is now clear why the fraction $1/\left( \Box - \frac{2}{3} {R} \right)$ is located on the right in the definition of $\gamma_{\rm C}(\Box)$ in (\ref{formADS}).
% The exponentials $\exp H_\ell$ ($\ell =0,2$) must be defined through the power series in
% $H_\ell$ that in turn are infinite series of polynomials.
%
%
This in turn results in that  the second variation of the action simplifies to,
 \be
 S^{(2)  {\rm CR}}_{\rm (A)dS} = \frac{1}{2} \int d^4 x \sqrt{|\bar{g}|} \left\{ \tilde{h}^{\bot \mu\nu}
\left( \Box - \frac{\bar{R}}{6} \right)
 \,  e^{H_8 \left( \Box + \frac{\bar{R}}{3} \right)} \,
\tilde{h}^{\bot}_{ \mu\nu}
- \tilde{\phi} \left( \Box + \frac{\bar{R}}{3} \right)  \,  e^{H_8 \left( \Box + \frac{\bar{R}}{3}\right)} \,  \tilde{\phi} \right\} ,
\label{variationADS3}
\ee
where $H_8$ is in the class of entire functions \eqref{TomboFF} with the polynomial $p(z)$ substituted by the above definition of the polynomial $p_8(\Box + \bar{R}/3)$ evaluated on the (A)dS background of curvature $\bar{R}$.

With a globally well defined field redefinition we can now completely remove the form factor and the
kinetic operator turns into the one of Einstein-Hilbert theory with cosmological constant.
The interactions will get modified by the field redefinition as well, but the Feynman diagrams will stay the same.
However, in doing so we do not really need to equate the form factors in front of different spin modes. Therefore, for the moment the choice (\ref{poly8}) is just to make the second order variation of the nonlocal theory as much similar as we can to the Einstein-Hilbert one. %more for the cosmetic beauty of the model.
Details about such a field redefinition are explained in Sec. V.

\subsubsection*{Analysis of the ``noncommutative" form factors}
%\footnote{
In this quite technical subsection we study some properties of the form factors that will turn out to be crucial in Sec. IV about quantum finiteness.
Let us remind that the exponentials of one, two, or multiple matrices are defined by means of power series, namely
 \be
 && e^X = \sum_{k=0}^{\infty} \frac{1}{k !} X^k \, , \quad
\nonumber\\
&&  e^{X+Y} = \sum_{k=0}^{\infty} \frac{1}{k !}
 \left(X+Y \right)^k \, , \quad  \nonumber \\
 &&
 e^{X_1+X_2 + X_3 + \ldots +X_N } =\sum_{k=0}^{\infty} \frac{1}{k !} \left(X_1+X_2 + X_3 + \ldots+ X_N \right)^k \, .
 \label{expO}
 \ee
 For the form factors defined in (\ref{formADS}) with polynomials (\ref{ps}) and (\ref{pc}) we can make explicit the above formula (\ref{expO}) as follows,
 \be
&& e^{H_{0,2}(\Box, R)} = \sum_{n=0}^{\infty} \frac{1}{n !} H(\Box, R)^n \nonumber \\
&&\hspace{1.55cm}
= \sum_{n=0}^{\infty} \frac{1}{n !} \left\{ \frac{1}{2} \left[ \Gamma \left(0, p_{\rm S, C}(\Box, R)^2 \right)+\gamma_E  + \log \left( p_{\rm S, C}(\Box, R)^2 \right) \right] \right\}^n \\
&& \hspace{1.55cm} =
\sum_{n=0}^\infty \frac{1}{n !} \left\{\sum_{m=1}^{\infty} (-1)^{m-1} \frac{1}{m !} \frac{p_{\rm S, C}(\Box, R)^{2 m}}{2 m} \right\}^n = \sum_{s=0}^\infty c_s \, p_{\rm S, C}(\Box, R)^{2 s} \, ,
\label{expH02}
 \ee
 where the coefficients $c_s$ are obtained by comparing the last two sums above. Notice that $p(\Box, R)$ certainly commutes with itself, but its arguments, namely $\Box$ and $R$, do not commute. Therefore, it is still easily possible to apply the structure of the vertex functions found in \cite{Tombo}
to the case of a binomial like %(\ref{ps}) and
 (\ref{pc}) because the form factor $\gamma_{\rm C}$ %and $\gamma_{\rm S}$
 is a function of only one polynomial, namely
\be
\gamma_{\rm C} = \sum^{\infty}_{s=0} \tilde{c}_{s}  \left( \Box - \frac{2}{3} R\right)^{8 s -1}
= \tilde{c}_1 \left( \Box - \frac{2}{3} R \right)^7 + \tilde{c}_3  \left( \Box - \frac{2}{3} R\right)^{23 }
+ \dots
  %\quad c_{s=0} =0
 \qquad
 {\rm or} \qquad
\gamma_{\rm C} = \sum^{\infty}_{r=0} a_{r}  \left( \Box - \frac{2}{3} R \right)^{r} \, ,
\label{semplice}
\ee
for a proper and fixed choice of the coefficients $a_r$ given the coefficients $\tilde{c}_s$. Now we can apply
the formula presented in \cite{Kuzmin} and rigorously proved in
\cite{Tombo, TomboStudent} to the operator $\tilde{\Box} = \Box - 2 R/3$.
In particular we can introduce the following notation,
\be
\gamma_{\rm C} = \sum_r^{\infty}  a_r (\Box_M + \mathcal{I})^r \, , \quad \mathcal{I} = \tilde{\Box}-\Box_M \, ,
\quad \Box_M = \eta^{\mu \nu}\partial_\mu \partial_\nu
\ee
We also remind that around fixed Minkowski background the form factor in momentum space is the Fourier transform of
\be
\gamma_{\rm C} = \sum^{\infty}_{r=0} a_{r}  \left( \Box_M  \right)^{r} =
  \frac{1}{2}  \frac{1}{  \Box_M }
\left( e^{H_0 \left( \Box_M \right) }  - 1    \right)  \, ,
\ee
while the gravitons vertices come only from the perturbative expansion of $\mathcal{I}$.

%appear just as variations of $\Box-2 R/3-\Box_{M}$ \cite{TomboStudent}. this is unclear \con

Less trivial is to apply the formula in \cite{TomboStudent} to $\gamma_{\rm S}$ that we can express as follows,
\be
&& \gamma_{\rm S} = - \frac{1}{6} \frac{1}{\Box+\frac{R}{3}} \sum_{s=1}^{\infty} c_s  \, p_{\rm S}^{2 s} (\Box,R)
=  - \frac{1}{6} \frac{1}{\Box+\frac{R}{3}} \sum_{s=1}^{\infty} c_s  \left( \frac{1}{\Lambda^8}  \left( \Box+ \frac{{R}}{3} \right)^2 \Box^2 \right)^{2 s} \nonumber \\
&& \hspace{0.45cm}
= - \frac{1}{6} \frac{1}{\Box+\frac{R}{3}} \sum_{s=1}^{\infty} c_s
\left( \frac{1}{\Lambda^8}  \left( \Box+ \frac{{R}}{3} \right)^2 \Box^2 \right)
\left( \frac{1}{\Lambda^8}  \left( \Box+ \frac{{R}}{3} \right)^2 \Box^2 \right)^{2 s-1}
\nonumber \\
&& \hspace{0.45cm}
= - \frac{1}{6} \left( \Box^3+\frac{R}{3}\Box^2 \right)  \sum_{s=1}^{\infty} \frac{c_s}{\Lambda^{16 s}}  \left(\left( \Box+ \frac{{R}}{3} \right)^2 \Box^2 \right)^{2 s-1} \!\!\!\!\! \nonumber \\
&&\hspace{0.45cm}
= - \frac{1}{6} \left( \Box^3+\frac{R}{3}\Box^2 \right)  \sum_{r=0}^{\infty} a_r \left(\left( \Box+ \frac{{R}}{3} \right)^2 \Box^2 \right)^{r}
\!\!  .
\label{piudiff}
\ee
The coefficients $c_s$ are fixed using the definition (\ref{expH02}), while the coefficients $a_r$
can be derived comparing the last two expressions in (\ref{piudiff}).
Now we can apply the derivation in \cite{TomboStudent} to
\be
\sum_{r=0}^{\infty} a_r \left( \Box_M^4 + \mathcal{I}  \right)^{r} \, , \quad  \mathcal{I} = \Box^4 - \Box_M^4 + O(R) \, .
\ee
When we expand in the graviton field we get interaction vertices coming from the variation of the binomial on the left of the sum in (\ref{piudiff}) and other vertices come from the variation of the sum. However, the full nonlocal contribution resulting from the variation of (\ref{piudiff}) will reconstruct
the same incremental ratios as defined in \cite{TomboStudent}, but for the form factor in Minkowski space rescaled by $1/\Box^4$, namely
\be
\frac{\gamma_{\rm S}(\Box_M)}{\Box_M^4} = - \frac{1}{6}   \frac{ e^{H_0 \left( \Box_M \right) %+ \frac{R}{3} \right)
}  - 1}{\Box_M^5}  \, \Lambda^8    \, .
\ee
%Therefore,
We can forget the nonlocality to evaluate the divergent contributions to the quantum effective action.% as well as for the theory without the Ricci scalar involved in the form factor \cite{modestoLeslaw}.
% or with a more symmetric structure like (\ref{semplice}).
%
%8 s -1 =r, s = 0 -> r = 7 , s=2 --> r = 17, etc
% when the polynomial is properly split in a free and an interacting part. For the polynomial (\ref{psimply}) we can write
% \be
% p\left( \Box+ \frac{{R}}{3} \right) = \frac{\Box_M^4}{\Lambda^8} + \frac{1}{\Lambda^8} \left(\Box^4 - \Box_M^4 + \Box \frac{R}{3} \Box^2 + \frac{R}{3} \Box^3 + \frac{R^2}{9} \Box^2 \right) =  \frac{1}{\Lambda^8} \left( \Box_M^4 + \mathcal{I}  \right) \, ,
% \ee
% where $\Box_M$ is the d'Alembertian operator in Minkowski spacetime and all the interactions are embodied in $\mathcal{I}$. Furthermore, this short analysis allows us to infer about the finiteness of the theory just considering the polynomial local theory in the UV regime.
  % and not in $\Box$ %in order to respect the ordering mentioned above
 %consistently with the property of $\gamma_{\rm C}$ and $\gamma_{\rm S}$ to be entire functions.
% This is equivalent to specify a particular ordering in the definition of the exponential entire functions and will be applied again later to the theory (\ref{TS}).

\subsection{A class of theories in the Ricci basis}
As a second example we consider the following action involving the Ricci tensor, but not the Weyl tensor in the quadratic in curvature part of the action, namely
\be
&& \boxed{\mathcal{L}_{\rm SR} = -  2 \kappa_4^{-2} \sqrt{|g|}\Big[ { R} -   2\Lambda_{\rm cc}
 %{\bf C} \, %\mu \nu \rho \sigma}
%  \gamma_{\rm C} (\Box) {\bf C}
  + {\bf S} \gamma_{\rm S2}(\Box) {\bf S }
  + { R}  \gamma_{\rm S}(\Box) { R }
  %C^{\mu \nu \rho \sigma}
  + {\cal V}({\bf C})
 \Big] } \,  ,
 \label{TS}
\ee
where the rank-two tensor ${\bf S}$ is defined by
\be
S_{\mu\nu} = R_{\mu\nu} - \frac{1}{4} R g_{\mu\nu}.
\label{Sdef}
\ee
In $D=4$ it is identically zero when evaluated on an (A)dS background and, moreover, it is completely trace-free.
%For the sake of simplicity this time we take $\gamma_{\rm C} = 0$ and we choose the following
The form factors in the action (\ref{TS}) are defined by:
\be
 &&
\boxed{\gamma_{\rm S2}(\Box) =    \frac{1}{  \Box - \frac{R}{6}   } \left(
e^{H_{\rm S2} \left( \left( \Box -  \frac{R}{6} \right) \left( \Box -  \frac{R}{3} \right) \right) } -1 \right)
}
  \, , \quad
  \\
&&
\boxed{
\gamma_{\rm S}(\Box) = - \frac{1}{6}  \frac{1}{ \Box+ \frac{R}{3} }
\left( e^{H_0 \left( \Box, R \right)
%+ \frac{R}{3} \right)
} -1 \right) %}{  \Box+ \frac{R}{3}  }
- \Box  \, \frac{1}{12  \left(  \Box + \frac{R}{2}  \right)  }
\left(
e^{H_{\rm S2} \left( \left( \Box +  \frac{R}{2} \right) \left( \Box +  \frac{R}{3} \right) \right) } -1 \right) \frac{1}{ \left( \Box + \frac{R}{3} \right)}
}
\, .
\label{formADSES}
\ee
We should here clarify how the entire functions defined above depend on their arguments.
Let us start with $H_{S2}$ in $\gamma_{\rm S2}(\Box)$, that is defined to be the following entire function of the polynomial $p_{\rm S2}$ of a fourth degree in $\Box$,
\be
&&
%\hspace{-0.5cm}
H_{\rm S2}  \left( %p_{\rm S2} (\Box, R )
\left( \Box-  \frac{R}{6} \right) \left( \Box -  \frac{R}{3} \right)
 \right)
   = \frac{1}{2} \left\{ \gamma_E
+ \Gamma \left( 0 ,  %\left[
 p^2_{\rm S2} (\Box, R ) %p\left( \left( \Box -  \frac{R}{6} \right) \left( \Box -  \frac{R}{3} \right) \right) \right]^2
\right)
+ \log \left(  p^2_{\rm S2} (\Box, R )
%p \left( \left( \Box -  \frac{R}{6} \right) \left( \Box -  \frac{R}{3} \right) \right)
\right) \right\}  ,
 \nonumber
 \\ &&
 %\hspace{-0.5cm} {\rm where} \quad
p_{\rm S2} (\Box, R ) % \! \left( \left( \Box -  \frac{R}{6} \right) \left( \Box -  \frac{R}{3} \right) \right) =
= \left( \Box -  \frac{R}{6} \right)^2 \left( \Box -  \frac{R}{3} \right)^2 \,  .
 \label{PolyHS2}
\ee
Therefore, the exponentiated entire function
$H_{\rm S2} \left( \left( \Box +  \frac{R}{3} \right) \left( \Box +  \frac{R}{2} \right) \right)$
in the second analytic operator in (\ref{formADSES})
is obtained translating the operator $\Box$ by the amount $\frac{2}{3} R$, namely
\be
H_{\rm S2} \left( \left( \Box +  \frac{R}{2} \right) \left( \Box +  \frac{R}{3} \right) \right) : =
H_{\rm S2} \!\left( \left( \Box -  \frac{R}{6} \right) \left( \Box -  \frac{R}{3} \right) \right)\Bigg|_{\Box \rightarrow \Box + \frac{2}{3} R} %\hspace{-0.5cm} .
.
\ee
Notice how the translated $\Box$ operators at the denominator in (\ref{formADSES})
have been placed in order to avoid ordering issues\footnote{
\label{f5} We can use here different definitions of the form factors to avoid the ordering problems of the denominators versus the exponential form factors, namely
\be
 &&
\gamma_{\rm S2}(\Box) =    \frac{
e^{H_{\rm S2} \left( \left( \Box -  \frac{4 \Lambda_{\rm cc} }{6} \right) \left( \Box -  \frac{4 \Lambda_{\rm cc}}{3} \right) \right) } -1
}{  \Box - \frac{4  \Lambda_{\rm cc}}{6}   }   \, , \quad
\label{formADSES2A}
\\
&&
\gamma_{\rm S}(\Box) = - \frac{1}{6}  \frac{ e^{H_0 \left( \Box+ \frac{4 \Lambda_{\rm cc}}{3} \right) } -1 }{ \Box+ \frac{4 \Lambda_{\rm cc}}{3} }
- \Box  \, \frac{
e^{H_{\rm S2} \left( \left( \Box +  \frac{4 \Lambda_{\rm cc}}{3} \right) \left( \Box +  \frac{4 \Lambda_{\rm cc}}{2} \right) \right) } -1   }{12 \left( \Box + \frac{4 \Lambda_{\rm cc}}{3} \right) \left(  \Box + \frac{4 \Lambda_{\rm cc}}{2}  \right)  }
\, .
\label{formADSES2B}
\ee
The ordering is now irrelevant, because $\Lambda_{\rm cc}$ is a numerical constant.
%Note that the rep
}.
%%%%
Moreover, taking the ``formal" limit $R \rightarrow 0$ in the form factors (\ref{formADSES}) and assuming $H_{\rm S2} = H_0$
the above %choice of the form factors turns the
Lagrangian (\ref{TS}) turns into %the action (\ref{EinsteinBasisS}).
\be
&& \mathcal{L}_{\rm E} = -  2 \kappa_D^{-2} \sqrt{|g|}\Big[ { R} +
 G_{\mu\nu}%\mu \nu \rho \sigma}
\,   \gamma_{\rm G} (\Box) R^{\mu\nu}
  %C^{\mu \nu \rho \sigma}
  + {\cal V}
 \Big]  \, , \nonumber \\
 &&
 %{\rm with} \quad
  \gamma_{\rm G} = \frac{e^{H_2} -1 }{\Box} \, .
\label{EinsteinBasisS}
\ee
Finally, the second order variation of the action for the Lagrangian (\ref{TS}) reads
\be
&& S^{(2)  {\rm CSR }}_{\rm (A)dS} = \frac{1}{2} \int d^4 x \sqrt{|\bar{g}|} \left\{ \tilde{h}^{\bot \mu\nu}
\left( \Box - \frac{\bar{R}}{6} \right) \left[ 1 + 2 \gamma_{\rm S}(0) \bar{R}
+  \left( \Box - \frac{\bar{R}}{6} \right) \gamma_{\rm S2} ( \Box)
\right] \tilde{h}^{\bot}_{ \mu\nu} \right. \nonumber \\
&& \left.
\hspace{4cm}
- \tilde{\phi} \left( \Box + \frac{\bar{R}}{3} \right) \left[ 1 + 2 \gamma_{\rm S}(0) \bar{R}
-  6 \left( \Box + \frac{\bar{R}}{3} \right) \, \gamma_{\rm S}( \Box ) - \frac{1}{2} \Box  \,
\gamma_{\rm S2} \left( \Box + \frac{2}{3} \bar{R} \right)
\right] \tilde{\phi} \right\} \, .
\label{variationADSES3}
\ee
We also selected out a form factor such that  $\gamma_{\rm S}(0) =0$.
For this purpose, after looking at the formula (\ref{formADSES}),
it is sufficient to take the following asymptotic polynomial $p_{\gamma +1}$
(for $\gamma+1 = 3+1$) as an argument of $H_0$,
\be
p_{\rm S}%{3 +1}
(\Box, R) = \left( \Box + \frac{R}{3} \right)^{2} \Box^2 .
\label{p31}
\ee
Therefore, after plugging
the form factors (\ref{formADSES}) in the second order variation (\ref{variationADSES3}) we end up again with (\ref{variationADS2}),
but with $H_2$ replaced by $H_{\rm S2}$, namely
 \be
 S^{(2)  {\rm CR}}_{\rm (A)dS} = \frac{1}{2} \int d^4 x \sqrt{|\bar{g}|} \left\{ \tilde{h}^{\bot \mu\nu}
\left( \Box - \frac{\bar{R}}{6} \right) \, e^{H_{\rm S2} \left(  \left( \Box - \frac{\bar{R}}{6} \right)
\left(  \Box - \frac{\bar{R}}{3}  \right) \right)}
\, \tilde{h}^{\bot}_{ \mu\nu}
- \tilde{\phi} \left( \Box + \frac{\bar{R}}{3} \right) \, e^{H_0 \left( \Box, \bar{R}  \right) } \, \tilde{\phi} \right\}.
\label{variationADS22}
 \ee
 %We now ...
In order to end up with the same form factor in the spin-two as long as in the spin-zero
graviton sectors we slightly modify the polynomial in (\ref{PolyHS2}) and we replace the curvature
$R$ with the cosmological constant $\Lambda_{\rm cc}$, namely
\be
\hspace{-0.2cm}
\tilde{p}(\Box; \Lambda_{\rm cc} )=  % \! \left( \left( \Box -  \frac{R}{6} \right) \left( \Box +  \frac{R}{3} \right) \left( \Box -  \frac{R}{3} \right) \Box \right) =
 \left( \Box -  \frac{R}{6} \right)^2 \left( \Box +  \frac{R}{3} \right)^2 \left( \Box -  \frac{R}{3} \right)^2
 \Big|_{ R \, \rightarrow \, 4 \Lambda_{\rm cc} } \!\!\!\!\! \times \Box^2 %\nonumber \\
 %&& \hspace{5.65cm}
 = \left( \Box -  \frac{2}{3}  \Lambda_{\rm cc} \right)^2 \left( \Box +  \frac{4}{3} \Lambda_{\rm cc} \right)^2 \left( \Box -  \frac{4}{3} \Lambda_{\rm cc} \right)^2 \Box^2 \, ,
 \label{pub}
 %\\
%&&
% = \tilde{p} \! \left( \left( \Box -  \frac{2}{3} \Lambda_{\rm cc} \right) \left( \Box +  \frac{4}{3} \Lambda_{\rm cc} \right) \left( \Box -  \frac{4}{3} \Lambda_{\rm cc} \right) \Box \right)
 % \, ,
\ee
where we technically replaced the Ricci scalar $R$ with $4 \Lambda_{\rm cc}$ to end up with a form factor without ordering issues. This replacement does not mean that we evaluate the form factor on the background, but just that the form factor has a particular (a posteriori) dependence on the constant $\Lambda_{\rm cc}$.
In (\ref{pub}) the  untranslated $\Box$ on the right secures that $\gamma_{\rm S}(0) =0$.
The form factors now read:
\be
 && \hspace{-1cm}
\gamma_{\rm S2}(\Box) =    \frac{
e^{H_{\rm S2} \left(  \tilde{p}  \left( \left( \Box -  \frac{2}{3} \Lambda_{\rm cc} \right) \left( \Box +  \frac{4}{3} \Lambda_{\rm cc} \right) \left( \Box -  \frac{4}{3} \Lambda_{\rm cc} \right) \right)\right) } -1 }{  \Box - \frac{2}{3}  \Lambda_{\rm cc}  }
%\Big|_{R \, \rightarrow \, 4 \Lambda_{\Lambda_{\rm cc}}}
  \, , \quad \\
&& \hspace{-1cm}
\gamma_{\rm S}(\Box) = - \frac{1}{6}  \frac{
 e^{H_{\rm S2}  \left( \tilde{p}  \left( \left( \Box -  \frac{2}{3} \Lambda_{\rm cc} \right) \left( \Box +  \frac{4}{3} \Lambda_{\rm cc} \right) \left( \Box -  \frac{4}{3} \Lambda_{\rm cc} \right) \right) \right) } -1 }{  \Box+ \frac{4}{3}  \Lambda_{\rm cc}}
- \frac{\Box}{12}  \,
\frac{
e^{H_{\rm S2} \left(  \tilde{p}  \left( \left( \Box -  \frac{2}{3} \Lambda_{\rm cc} \right) \left( \Box +  \frac{4}{3} \Lambda_{\rm cc} \right) \left( \Box -  \frac{4}{3} \Lambda_{\rm cc} \right) \right)\right)\big|_{\Box \, \rightarrow \Box + \frac{8}{3} \Lambda_{\rm cc} } } -1 }{ \left( \Box + 2 \Lambda_{\rm cc}  \right)\left( \Box + \frac{4}{3} \Lambda_{\rm cc} \right)}
\, .
\label{formADSESconst}
\ee
Therefore, we end up with the form factors already introduced in the footnote above, but with a new polynomial
as an argument of the entire function $H_{\rm S2}$. Moreover, now $H_0 = H_{\rm S2}$ and the second variation of the action (\ref{variationADS22}) turns into
 \be
 S^{(2)  {\rm CR}}_{\rm (A)dS} = \frac{1}{2} \int d^4 x \sqrt{|\bar{g}|} \left\{ \tilde{h}^{\bot \mu\nu}
\left( \Box - \frac{\bar{R}}{6} \right) \, e^{H_{\rm S2} \left( \tilde{p} \right)}
\, \tilde{h}^{\bot}_{ \mu\nu}
- \tilde{\phi} \left( \Box + \frac{\bar{R}}{3} \right) \, e^{H_{\rm S2} \left( \tilde{p} \right) } \, \tilde{\phi} \right\}.
\label{variationADS23}
\ee
The second order variation (\ref{variationADS23})
has the same form factor to multiply the tensorial as long as the scalar perturbations.
%This subclass of theories

Once more we point out that the replacement of $R$ with $\Lambda_{\rm cc}$ is an off-shell operation just as in Einstein-Hilbert theory in the presence of a cosmological constant.

%\section{Einstein similar action with cosmological constant}
\section{Quantum Finiteness} \label{s4} %nonlocal gravity in (A)dS vacuum} %on MSS background}
%As particular case we can assume $H_2(\Lambda_{\rm cc} = 0) = H_0(\Lambda_{\rm cc} = 0)$.
%
%In this section we prove that it is possible
In this section we study two classes of theories involving respectively the Ricci scalar and the off-shell cosmological constant in the form factors. In the first subsection we study the theory  (\ref{TWeyl3}) with form factors (\ref{formADS}), while in the second subsection the theory (\ref{TWeyl3}) with form factors (\ref{formADS2}).

\subsection{Analysis of the theory (\ref{TWeyl3}) with form factors (\ref{formADS})
} \label{FiniteADS}
 %In this section we provide at least an example of theory finite at quantum level.
In agreement with the analysis in the previous section the polynomial appearing in the
ultraviolet limit of the form factor can also contain powers of the Ricci scalar, while the nonlocal structure only gives contributions to the finite part of the quantum effective action.
Therefore, a quite general polynomial giving a contribution to the beta functions in $D=4$ is:
%assuming the polynomial $p(z)$ in (\ref{Tomboulis}) to be at least a trinomial, namely
\be
&& \hspace{-0.5cm}
p(z, \mathcal{R} ) = a^{(0)}_{\gamma+1} z^{\gamma+1}
+ a^{(1)}_{\gamma+1} z^{\gamma} \, \mathcal{R}
+ a^{(2)}_{\gamma+1} z^{\gamma-1} \mathcal{R}^2 + \ldots  \nonumber \\
&& \hspace{1cm}
+  a^{(0)}_{\gamma} z^{\gamma} + a^{(1)}_{\gamma} z^{\gamma-1} \, \mathcal{R}
%\cor\cancel{+a^{(2)}_{\gamma-2} z^{\gamma-2} \mathcal{R}^2} \con
+ \ldots \nonumber \\
&& \hspace{1cm}
 +  a^{(0)}_{\gamma-1} z^{\gamma-1} \, .
%\cor\cancel{+ a^{(1)}_{\gamma-1} z^{\gamma-2} \, \mathcal{R}+ a^{(2)}_{\gamma-1} z^{\gamma-3} \, \mathcal{R}^2} ???\con + \ldots
\label{polyn}
\ee
%where here $z= (\Box - \frac{2}{3} \Lambda_{\rm cc})/\Lambda^2$ or $z= (\Box + 2 \Lambda_{\rm cc})/\Lambda^2$ for $H_2$ and $H_0$ respectively,
%then the theory likely turns out to be finite.
%%
%%
%%
For the theories presented in this paper $\mathcal{R}$ can only be the Ricci scalar.
The ellipses ($\ldots$) also include terms arising from commutators of the $\Box$ operator
 with covariant derivatives and curvatures.

To have a finite theory at quantum level (or better conformally invariant) we have to make vanish the beta functions
for the following six operators,
\be
\sqrt{|g|} \, , \quad {\sqrt{|g|} R} \, , \quad
\sqrt{|g|} { R}^2 \,, \quad
\sqrt{|g|}{\bf Ric}^2 \, , \quad
\sqrt{|g|} {\bf GB} \, , \quad
\sqrt{|g|} \Box R \, ,
\label{divbasis}
\ee
where {\bf GB} is the Gauss-Bonnet operator.
The beta functions $\beta_{R^2}$, $\beta_{R_{\mu \nu}^2}$ and $\beta_{\bf GB}$ (in the basis \eqref{divbasis})
 get contributions also from the following
killers, if they are added to the action,% (\ref{killersW}), namely
\be
{\cal V}({\bf C}) = s^{(1)}_w \, C_{\mu\nu\rho\sigma} C^{\mu\nu\rho\sigma} \Box ^{\gamma -2} C_{\alpha \beta \gamma \delta}
C^{\alpha \beta \gamma \delta} +
s^{(2)}_w \, C_{\mu\nu\rho\sigma} C^{\alpha \beta \gamma \delta} \Box ^{\gamma -2} C_{\alpha \beta \gamma \delta}
C^{\mu\nu\rho\sigma} \,.
\label{killersW}
\ee
These killers do not spoil the structure of the kinetic operator nor the propagator because
%are compatible with
%the result (\ref{withCC3}) because
the Weyl tensor evaluated on any homogeneous and isotropic spacetime is identically zero and the second order variation of the action based on (\ref{killersW}) is at least quadratic in the Weyl tensor.
Moreover, they are enough to make zero the two beta functions $\beta_{R^2}$ and
$\beta_{R_{\mu \nu}^2}$. Indeed, the contribution of (\ref{killersW}) can only be linear in the front coefficients $s^{(1)}_w$ and $s^{(2)}_w$ as has been shown in \cite{modestoLeslaw} by
a direct implementation of the background field method.

If we want to use killers that do not change the structure of the kinetic operator around (A)dS one option is to build them using only hatted quantities like in the footnote \ref{f5} (so with the background value of the tensor subtracted, cf. also Appendix). Other viable killers, which possess the same property, are:
\be
s_s^{(1)} \, {S}_{\mu \nu} {S}^{\mu \nu} \Box^{\gamma - 2}
{S}_{\rho \sigma} {S}^{\rho \sigma}
\, , \quad
s_s^{(2)} \, {S}_{\mu \nu} {S}^{\rho \sigma} \Box^{\gamma - 2}
{S}_{\rho \sigma} {S}^{\mu \nu} \,  , \quad \mbox{where} \quad
S_{\mu\nu} \quad \mbox{ was defined in \eqref{Sdef}} \, .
\label{killersS}
\ee
%The above killers turn out to be \cor crucial ? why \con for cancellation of the divergences proportional to the {\bf GB} operator and the \cog $\Box R$ operator that are not trivially zero in (A)dS. \con
%\cor I do not agree!
On any MSS background the $\bf GB$ operator is nonvanishing, while $\Box R$ always vanishes. Regarding the contributions to the divergent part of the quantum effective action (under an integral) $\bf GB$ and $\Box R$ can be neglected as total derivatives on MSS. %when suitable boundary conditions are employed.
The reason to kill these two more divergences has eventually to do with the conformal invariance of the theory, but not merely with finiteness.

The beta function for the Newton constant $\beta_{R}$ can be made zero using following example of
the killer operator
\footnote{Additionally, we can make to vanish the beta functions
$\beta_{R^2}$, $\beta_{R_{\mu \nu}^2}$  and $\beta_{R}$  introducing also the following terms in $\cal V$,
\begin{eqnarray}
&& s_r^{(1)} \,  \hat{R}^2\Box^{\gamma -2}
 \hat{R}^2, \,   \quad s_r^{(2)} \, \hat{R}_{\mu \nu} \hat{R}^{\mu \nu} \Box^{\gamma - 2}
\hat{R}_{\rho \sigma} \hat{R}^{\rho \sigma}\quad  {\rm and}  \quad
 s_r^{(3)}\hat{R}_{\mu \nu} \hat{R}^{\mu \nu} \Box^{\gamma - 2}
\hat{R}  \quad {\rm respectively,} \label{killersHat}\\
 && \mbox{where} \quad
 \hat{R}_{\mu\nu} = R_{\mu\nu} - \Lambda_{\rm cc} g_{\mu\nu}\quad{\rm and}\quad\hat{R}=R-4\Lambda_{\rm cc}\,. \nonumber
\end{eqnarray}
However, the operators (\ref{killersHat}) must be used more carefully because the cosmological constant can be present in the beta functions.},
\be
S_{\mu\nu} S^{\mu}_\rho \Box^{\gamma -2} S^{\nu\!\rho} \, .
\ee
Finally, to have a finite theory we need to make vanishing the beta function for the cosmological constant.
For this achievement we need to explicitly evaluate the divergent contributions to the one-loop
effective action that do not contain any curvature tensor.
This result was derived for the first time in \cite{shapiro3} and also
successfully attained by our group \cite{ShapiroLeslawModesto}.
Given the polynomial (\ref{polyn}), only the monomials independent on the curvatures can
contribute to the $\mathcal{R}^0$ divergence. Therefore, the beta function can only depend on the
coefficients $a^{(0)}_{\gamma+1}, a^{(0)}_{\gamma}, a^{(0)}_{\gamma-1}$ in (\ref{polyn}).
For the sake of simplicity we here only consider
the theory in Weyl basis (\ref{TWeyl3})
with form factors (\ref{formADS}). Moreover, we take $H_2=H_0$, but we replace the polynomial (\ref{poly8}) with
\be
 {p}_{12} = \left( \Box+ \frac{R}{3} \right)^2
  \left( %a^{(0)}_{\gamma+1}
  c_1
  \Box^3
  +
   %a^{(0)}_{\gamma}
   c_2
   \Box^2
+
 %a^{(0)}_{\gamma-1}
 c_3
 \Box\right)^2
  \left( \Box - \frac{2}{3} R \right)^4
 =   a^{(0)}_{\gamma+1} \Box^{12}
  +    a^{(0)}_{\gamma} \Box^{11}
+ a^{(0)}_{\gamma-1} \Box^{10} + O(R) \, ,
\label{poly12}
\ee
and comparing with (\ref{polyn}): $a^{(0)}_{\gamma+1} = c_1$, $a^{(0)}_{\gamma}=2 c_1  c_2$, $a^{(0)}_{\gamma-1} = 2 c_1 c_3$.
Note that with the polynomial (\ref{poly12}) we surely avoid the issue of nonlocal counterterms because it is definite positive on the real axis (namely $\sqrt{{p}_{12}^2} = {p}_{12}$). Therefore,
$c_1$, $c_2$ and $c_3$ can be selected to be positive, negative or zero (at least one of the $c_i$ must be nonzero).

The form factors $\gamma_{\rm C}$ and $\gamma_{\rm S}$ in the UV, are respectively,
\be
&&
\gamma_{\rm C} \rightarrow \frac{e^{\gamma_E/2} }{2} \left( \Box+ \frac{R}{3} \right)^2
  \left( %a^{(0)}_{\gamma+1}
  c_1 \Box^3 +   c_2  \Box^2 + c_3  \Box\right)^2
  \left( \Box - \frac{2}{3} R \right)^4 \, , \label{asiFormC}\\
  &&
  \gamma_{\rm S} \rightarrow - \frac{e^{\gamma_E/2} }{6} \left( \Box+ \frac{R}{3} \right)
  \left( %a^{(0)}_{\gamma+1}
  c_1 \Box^3 +   c_2  \Box^2 + c_3  \Box\right)^2
  \left( \Box - \frac{2}{3} R \right)^4 \, .
  \label{asiForm}
\ee
Moreover, the operators $O(R)$ do not give contribution to the beta function for the cosmological constant.
Finally, we need to explicitly evaluate the beta function for the cosmological constant
$(\beta_{\Lambda_{\rm cc}}$) and select the parameters $a^{(0)}_{\gamma+1}, a^{(0)}_{\gamma}, a^{(0)}_{\gamma-1}$ to make zero $\beta_{\Lambda_{\rm cc}}$. Once more, the parameters
$a^{(0)}_{\gamma+1}, a^{(0)}_{\gamma}, a^{(0)}_{\gamma-1}$ do not run because all of them appear in front of higher derivative operators of dimension higher than four.

For the theory (\ref{TWeyl3}) with form factors (\ref{formADS}) we can explicitly show the finiteness of the theory because the beta function $\beta_{\Lambda_{\rm cc}}$ has been computed in \cite{shapiro3,ShapiroLeslawModesto} for the following prototype theory
%(see footnote four for more details and issues about the asymptotic behavior of the theory),
%
%For this purpose we only need to study the following prototype local theory,
\be
S_{N} =
\int \!d^4x \sqrt{|g|}\, \left(
\omega_{N,R} R \, \Box^{N} R
+ \omega_{N,C} C \, \Box^{N} C \right) \, .
\label{CCdivsaction}
\ee
From the divergent part of the quantum effective action we can read the beta function.
The outcome of the computation is \cite{shapiro3}:
%and the beta function for the cosmological constant reads
\be
\Gamma^{(1)}_{\rm cc,\,div}
= -
\frac{1}{2 (4 \pi)^2} \frac{1}{\epsilon}  \int \!d^4 x
\sqrt{|g|}
\left(\frac{5 \omega _{N-2,C}}{\omega _{N,C}}+\frac{\omega _{N-2,R}}{\omega _{N,R}}-\frac{5 \omega _{N-1,C}^2}{2 \omega _{N,C}^2}-\frac{\omega _{N-1,R}^2}{2 \omega
   _{N,R}^2}\right)
 %\hspace{1.13cm}
    \equiv -
\frac{1}{2 \epsilon} \! \int \!d^4 x
\sqrt{|g|}\, \beta_{\Lambda_{\rm cc}}
   \, .
\label{CCdivs}
\ee
Finally, we have to compare the nonrunning coefficients
$\omega_{i,{C}}$ and $\omega_{i,{R}}$ ($i=N+1, \, N,  \, N -1$), which appear in front of the operators quadratic in the Weyl tensor and in the Ricci scalar in (\ref{CCdivsaction}), with the parameters in front of the same operators resulting in the action (\ref{TWeyl3}) with asymptotic form factors (\ref{asiFormC}) and (\ref{asiForm}).

 Since the issue with UV divergences is probing the UV limit of the theory this can be also thought in the following way. The divergences arise because of the coincidence limit of points used as arguments of Green's  functions. When points do come closer the spacetime is effectively flat and they do not see such effect like the (A)dS curvature radius. That is why all divergences on MSS are the same as on the flat spacetime. Finally, the UV divergences in QFT do not depend on the background and, therefore, we have background independence of superrenormalizability or finiteness. %The same is with the ways to make it UV finite.
In other words, if the theory is UV finite around the flat spacetime, then it is also finite around any other background, in particular this applies to MSS backgrounds.
%{\color{blue}Moreover, for an on-shell background $\Lambda_{\rm cc}$ is just a number so that the dependence on it in the adjusted coefficients of killers $s_i$ is immaterial and it can enter as a rational function. Therefore, there is no a problem in formulating
%defining
%for the formulation of a Lagrangian for
%a theory with $\Lambda_{\rm cc}$}. \con
%{\color{red}We here assumed that the formula in \cite{Tombo, TomboStudent} or a generalization of that exists and can be applied to a theory with a general asymptotic polynomial involving the Ricci scalar.}

\subsection{Analysis of the theory (\ref{TWeyl3}) with form factors (\ref{formADS2})}
%We now make a remark about the form factors (\ref{formADS2}) and (\ref{formADSES2A}), (\ref{formADSES2}).
We hereby consider the theory (\ref{TWeyl3}) with form factors (\ref{formADS2}).
These form factors [and also (\ref{formADSES2A}), (\ref{formADSES2B})] depend explicitly on the cosmological constant
$\Lambda_{\rm cc}$ that in general could appear in the beta functions making the search for a finite quantum gravity much more involved.
However, it is sufficient to select out polynomials that in the UV regime do not involve the cosmological constant at least in the coefficients $\omega_{i,{C}}$ and $\omega_{i,{R}}$ for $i=N+1, \, N,  \, N -1$. Given the theory (\ref{TWeyl3}) with form factors (\ref{formADS2}) we can select the %take $H_0=H_2$ and the
following asymptotic polynomials,
\be
&&
p_{\rm C}\left( \Box; \Lambda_{\rm cc} % - \frac{8}{3} \Lambda_{\rm cc}
 \right) = \frac{1}{\Lambda^8} \Box^2 \left( \Box-  \frac{8}{3}  \Lambda_{\rm cc} \right)^2
 \left( \Box^2 + \frac{8 }{3} \Lambda_{\rm cc} \Box +  \left( \frac{8 }{3} \Lambda_{\rm cc}\right)^2 \right)
 \, , \label{ps3a}\\
&&
p_{\rm S}\left( \Box; \Lambda_{\rm cc}
%\left( \Box+ \frac{{4}}{3} \Lambda_{\rm cc}
 \right) = \frac{1}{\Lambda^8}  \Box^2 \left( \Box+ \frac{{4}}{3} \Lambda_{\rm cc} \right)^2
 \left( \Box^2 - \frac{4 }{3} \Lambda_{\rm cc} \Box +  \left( \frac{4 }{3} \Lambda_{\rm cc}\right)^2 \right)
 \, .
\label{ps3b}
\ee
%
%\be
%&& \hspace{-0.6cm}
%\tilde{p}_{18} = \frac{\Box^2}{\Lambda^{36}} \left( \Box+ \frac{4}{3} \Lambda_{\rm cc}\right)^2
%\left(  \Box^2 - \frac{4}{3} \Lambda_{\rm cc} \Box + \left( \frac{4}{3} \Lambda_{\rm cc} \right)^2 \right)^2
% \left(  \Box^2 + \frac{8}{3} \Lambda_{\rm cc} \Box + \left( \frac{8}{3} \Lambda_{\rm cc} \right)^2 \right)^4
%  \left( \Box - \frac{8}{3} \Lambda_{\rm cc} \right)^4 \nonumber \\
%  &&  \hspace{0cm} = \frac{\Box^2}{\Lambda^{36}}
%  \frac{\left(512 \Lambda_{\rm cc}^3-27 \Box^3 \right)^4 \left(64 \Lambda_{\rm cc}^3+27 \Box^3\right)^2}{387420489}
%= \frac{\Box^2}{\Lambda^{36}}
% \left[\frac{281474976710656 \Lambda_{\rm cc}^{18}}{387420489} + \right.
%   \nonumber \\
%   && \left.
%    \hspace{0cm}
%   +\frac{2199023255552 \Lambda_{\rm cc}^{15}
%   \Box^3}{4782969}+\frac{2147483648 \Lambda_{\rm cc}^{12} \Box^6}{177147}
%   -\frac{343932928 \Lambda_{\rm cc}^9
%   \Box^9}{19683}+\frac{438272 \Lambda_{\rm cc}^6 \Box^{12}}{243}-\frac{640 \Lambda_{\rm cc}^3 \Box^{15}}{9}+\Box^{18} \right]  .
%\label{poly18tilde}
%\ee
Notice that the two parabolic trinomials on the right sides in (\ref{ps3a}) and (\ref{ps3b}) are positive for any value of $\Box$ and $\Lambda_{\rm cc} >0$.
For the above selected polynomials (\ref{ps3a}) and (\ref{ps3b}), $\omega_{N-1,C(R)}=0$ and $\omega_{N-2,C(R)}=0$.
Indeed,
\be
&&
\hspace{-0.6cm}
\gamma_{\rm C}(\Box) =   \frac{1}{2}  \frac{\left( e^{H_2 \left( \Box -  \frac{8}{3} \Lambda_{\rm cc}  \right) } -1 \right) }{\Box - \frac{8}{3}  \Lambda_{\rm cc} }
 \, \rightarrow \,  \frac{1}{2} \frac{1}{\Lambda^8}  \Box^2 \left( \Box - \frac{{8}}{3} \Lambda_{\rm cc} \right)
 \left( \Box^2 + \frac{8}{3} \Lambda_{\rm cc} \Box +  \left( \frac{8}{3} \Lambda_{\rm cc}\right)^2 \right)
 = \frac{1}{2 \Lambda^8} \left( \Box^5 - \frac{512 \Lambda_{\rm cc}^3  \Box^2}{27} \right)
 \, ,
\nonumber \\
&&\hspace{-0.6cm}
\gamma_{\rm S}(\Box) = - \frac{1}{6}   \frac{\left( e^{H_0 \left( \Box, \Lambda_{\rm cc} \right)
}  - 1    \right)}{\Box+ \frac{4}{3}  \Lambda_{\rm cc}} \, \rightarrow \, - \frac{1}{6} \frac{1}{\Lambda^8}  \Box^2 \left( \Box+ \frac{{4}}{3} \Lambda_{\rm cc} \right)
 \left( \Box^2 - \frac{4 }{3} \Lambda_{\rm cc} \Box +  \left( \frac{4 }{3} \Lambda_{\rm cc}\right)^2 \right)
 = -\frac{1}{6 \Lambda^8} \left( \Box^5 + \frac{64 \Lambda_{\rm cc}^3  \Box^2}{27} \right) \, .
 \nonumber
\ee
Therefore,
there is no contribution to the beta functions $\beta_{\Lambda_{\rm cc}}$ and $\beta_{\kappa}$. More importantly, the cosmological constant does not appear in any beta function. In general, we only need the beta function for the cosmological constant to be independent on $\Lambda_{\rm cc}$ itself to achieve  one-loop exact superrenormalizability or finiteness because $\kappa_4$ does not appear in the form factors and, therefore, in the beta functions. Let us expand a little on this point. If the beta functions do not depend on any of the running couplings then we can make them zero at any energy scale and at any loop order by adding suitably selected killer operators because the superrenormalizability implies that the beta functions are one-loop exact.

%The last example of theory is surely finite at quantum level, while it is needed a more careful analysis of the theory presented in the previous subsection because the form factor involves the Ricci scalar. % in the polynomial at the exponent.
%t the moment the analysis in this section applies only to the form factors independent on
%$\Lambda_{\rm cc}$.

%Moreover, we can choose the coefficients $\omega_{\gamma+1}$,
%$\omega_{\gamma}$ and $\omega_{\gamma-1}$ to make zero the beta function for $G_N$ and
%$\Lambda_{\rm cc}$. The structure of the beta functions is indeed ...
\section{Field redefinition \& tree-level perturbative triviality}\label{GFR}
In this section we explicitly show that for a large class of theories involving neither the Riemann nor the Weyl tensor, a field redefinition theorem provides an explanation for the stability of MSS
in weakly nonlocal theories. Namely all these theories are tree-level equivalent to the Einstein-Hilbert
theory in the presence of cosmological constant.
Let us consider the theory (\ref{TS}) with $\gamma_{\rm C}=0$, namely
%%%%%%%
%%%%%%%
\begin{comment}
\footnote{
Another theory with similar properties is:
\be
S_{{\rm NL}\Lambda} = -  2 \kappa_D^{-2} \int d^D x \sqrt{|g|}\Big[ {\bf R} - 2 \Lambda_{\rm cc} +
 (G_{\mu\nu}  + \Lambda_{\rm cc} g_{\mu\nu} ) \,
  \gamma_{\rm G} (\Box, {\bf Ric} , {\bf R} ) (R^{\mu\nu} - \Lambda_{\rm cc} g^{\mu\nu} )
  + {\cal V( \Box, {\bf Ric} , {\bf R} )}
 \Big]  \,  .
% \quad \gamma_{\rm G} = \frac{e^{H_2} -1 }{\Box} \, .
\label{EinsteinBasisFR}
\ee
The reason to introduce the translation of $\Lambda_{\rm cc} g_{\mu\nu}$ in the nonlocal sector of the theory will be clear shortly, however, their impact is harmless because their contribution is just a total derivative.
}
\end{comment}
%%%%%
%%%%%
\be
 S_{{\rm NL}-\Lambda}  = -  2 \kappa_4^{-2} \!\int\! d^4 x \sqrt{|g|}\Big[  R - 2 \Lambda_{\rm cc}
  + {\bf S} \gamma_{\rm S2}(\Box) {\bf S }
  + { R}  \gamma_{\rm S}(\Box) { R }
  %C^{\mu \nu \rho \sigma}
  + {\cal V}( \Box, {\bf Ric} ,  R )
  %{\cal V}({\bf C})
 \Big]  \, .
 \label{TS2}
\ee
%%%
We can now recast the above action in a way that explicitly shows the Einstein's gravitational EOM in the presence of a cosmological constant, i.e.
\be
%&&
E_{\mu\nu} = G_{\mu\nu} + \Lambda_{\rm cc} g_{\mu\nu} \, ,  \quad
%%%
%R_{\mu\nu}= E_{\mu\nu} - \frac{1}{2} g_{\mu\nu} E^\alpha_\alpha  + \Lambda_{\rm cc} g_{\mu\nu} \, ,
%\\
%&&
R_{\mu\nu} = E_{\mu\nu} - \frac{1}{2} g_{\mu\nu} E^\alpha_\alpha + \Lambda_{\rm cc} g_{\mu\nu} \, , \quad S_{\mu\nu} = E_{\mu\nu} - \frac{1}{4} g_{\mu\nu} E^\alpha_\alpha \, ,  \quad R = - E^\alpha_\alpha + 4 \Lambda_{\rm cc} \, .
\label{EHeom}
\ee
Making use of the EOM (\ref{EHeom}), the action now equivalently turns into
\be
&& S_{{\rm NL}-\Lambda} = -  2 \kappa_4^{-2} \!\int\! d^4 x \sqrt{|g|} \Big[  R - 2 \Lambda_{\rm cc}
+  \left( E_{\mu\nu} - \frac{1}{4} E g_{\mu\nu}  \right) \,
  \gamma_{\rm S2} (\Box, E, \Lambda_{\rm cc} )  \left( E^{\mu\nu} - \frac{1}{4} E g^{\mu\nu}  \right)
\nonumber  \\
&&
\hspace{4cm}
  +
  \left( E - 4 \Lambda_{\rm cc}   \right) \,
  \gamma_{\rm S} (\Box, E, \Lambda_{\rm cc} ) \left( E - 4 \Lambda_{\rm cc}   \right)
    + {\cal V}( \Box, {\bf E},  E, \Lambda_{\rm cc})
 \Big],
% \quad \gamma_{\rm G} = \frac{e^{H_2} -1 }{\Box} \, .
\label{EinsteinBasisFR2}
\ee
where ${\bf E}$ stays for $E_{\mu\nu}$ and $E\equiv E^\mu_\mu$.
The nonlocal form factor $\gamma_{\rm S}$ satisfies the property $\gamma_{\rm S}(0) =0$
[see, for example, (\ref{formADSESconst}) with the polynomial (\ref{pub})]. Therefore,
%term involving explicitly the cosmological constant in the curvature part, namely the term with $\gamma_{\rm S}$ form factor, \cor only one $\gamma_{\rm S}$, what with $\gamma_{\rm S2}???$ \con is selected in such a way in order to give zero or a total derivative contribution to the action
%(one example is given by $H_0$ with argument (\ref{p31})) and then
we can rewrite the action in the following simplified form,
\be
&&
S_{{\rm NL}-\Lambda} = -  2 \kappa_4^{-2} \!\int\! d^4 x \sqrt{|g|}\Big[ R - 2 \Lambda_{\rm cc} +
 E_{\mu\nu} \, F^{\mu\nu, \rho \sigma} \, E_{\rho \sigma}
 % \gamma_{\rm G} (\Box, {\bf E} ) (E^{\mu\nu}
  %- \frac{1}{2} E^\alpha_\alpha g^{\mu\nu} - 2 \Lambda_{\rm cc} g^{\mu\nu} )
  %+ {\cal V( \Box, {\bf E}  )}
 \Big]  \,  , \label{EinsteinBasisFR3}
  \\ &&
  F^{\mu\nu, \rho \sigma} \equiv
  \gamma_{\rm S2} (\Box, E, \Lambda_{\rm cc} ) \left(
 g^{\mu\rho} g^{\nu \sigma} -\frac{1}{4}  g^{\mu\nu} g^{\rho \sigma}  \right)
 +
  \gamma_{\rm S} (\Box, E, \Lambda_{\rm cc} ) \,   g^{\mu\nu} g^{\rho \sigma}
 +
  {\cal \tilde{V}}^{\mu\nu\rho\sigma}( \Box, {\bf E}, E, \Lambda_{\rm cc}  )
  \, ,
\label{FEinsteinBasisFR3}
\ee
where the potential ${\cal V}( \Box, {\bf E}, E, \Lambda_{\rm cc}  )$ must be at least quadratic in the EOM
$E_{\mu\nu}$, namely
\be
{\cal V}( \Box, {\bf E}, E, \Lambda_{\rm cc}  ) = E_{\mu\nu} {\cal \tilde{V}}^{\mu\nu\rho\sigma}( \Box, {\bf E}, E, \Lambda_{\rm cc}  ) E_{\rho\sigma}\, .
\label{termsVL}
\ee
%\cor Is it necessary that $\gamma_{\rm S}(0)=0$ to avoid cc terms? \con
In the view of the restructured
action (\ref{EinsteinBasisFR3}), we are now ready to implement the following
general theorem in the presence of a cosmological constant. % and based on the theorem proved in flat sp
An analogous theorem was previously proved and applied to the case without cosmological constant \cite{amplitudes}.
%%%%%%

{{\bf Theorem.}} {
By making use of a proper analytic field redefinition $g\to g'$
the action (\ref{EinsteinBasisFR3}) can be recast in
%All the $n$-point functions in gravitational theories
%$($in particular superrenormalizable or finite$)$ with action (\ref{EinsteinBasisFR3})
%give the same on-shell tree-level amplitudes as
the Einstein-Hilbert form with the presence of a cosmological constant term, i.e.
\be
 \mathcal{L}_{{\rm EH}-\Lambda }
 = -  2 \kappa_{4}^{-2} \, \sqrt{|g|} \left(  {  R } - 2 \Lambda_{\rm cc}
 \right),
 \label{EHLambda}
 \ee
 %by using a proper analytic field redefinition and
provided that ${\cal V}$ has the structure given in (\ref{termsVL}), namely
it is at least quadratic in $E_{\mu\nu}$ and/or $E^\alpha_\alpha$ and does not contain any Riemann or Weyl tensor explicitly. Therefore, the theorem does not apply to the theory with $\gamma_{\rm C} \neq 0$.

{{\bf Proof.}} %The proof is based on the field redefinition theorem proved in \cite{Anselmi:2006yh}
%at perturbative level and to all orders in the Taylor expansion of the redefinition of the metric field.
The proof is based on a perturbative field redefinition $g\to g'$ to all orders in the Taylor expansion with respect to the redefinition of the metric field.
First, we assume that we have given two general weakly nonlocal action functionals
$S'(g)$  and $S(g')$, respectively defined in terms of the metric fields
$g$ and $g'$, such that
\be
S'(g) =
%S(g')
 S(g) + E_i(g) F_{i j}(g)  E_j (g) \, ,
 \label{AnselmiC}
\ee
where $F$ can contain derivative operators and ${E_i = \delta S/\delta g_i}$ are the EOM of the theory with the action $S(g)$ \footnote{Here we use a compact deWitt notation and with the indices $i$, $j$ on fields we encode all Lorentz, group indices, and the spacetime dependence of the fields. Additionally, we assume that the field space is flat and we do not need to raise indices in sums there.}. The statement of the theorem is that there exists a field redefinition
\be
g_i'  = g_i + \Delta_{i j} E_j  \quad \Delta_{i j} = \Delta_{j\hspace{0.03cm} i},
\label{FRe}
\ee
such that, perturbatively in $F$, but to all orders in powers of $F$ in the field redefinition $g\to g'$ \,\footnote{The field redefinition preserves general covariance. Indeed, formula (\ref{AnselmiC}) shows a multiple product of weakly nonlocal factors $F _{i j}$ and the EOM $E_{i}$, which are both covariant under active general coordinate transformations (diffeomorphisms). Additionally, the asymptotic behavior of the field redefinitions is such that they go to zero sufficiently fast at infinity together with the fast falloff of fields in order to preserve the spectrum of the theory. As a corollary, the large diffeomorphisms, are not touched by such field redefinitions.}, we have the equivalence
\be
S'(g) =
S(g')  \,.
\label{FR}
%= S(g) + E_i F^{i j}  E_j (g),
\ee
Above $\Delta_{ij}$ is a possibly nonlocal operator acting linearly on the EOM $E_j$, with indices $i$ and $j$ in the field space, and it is defined perturbatively %in terms of
in powers of the operator $F_{ij}(g)$, namely $\Delta_{ij}=F_{ij}(g)+\ldots\,\,$.
% starting at, or infinitesimally
Let us consider the
first order in Taylor expansion for the functional $S(g')$, which reads
\be
S(g') = S(g + \delta g) \approx S(g) + \hspace{-0.05cm}  %\overbrace{
\frac{\delta S}{\delta g_i} %}^{ {\rm EOM} }
\hspace{-0.05cm}   \delta g_i =
 S(g) + E_i   \, \delta g_i \, .
\ee
If we can find a weakly nonlocal expression for $\delta g_i$ such that
%\be
$S'(g) = %S(g') \approx
S(g) + E_i   \, \delta g_i$
%\ee
(note that the argument of the functionals $S'$ and $S$ is now the same),
then there exists a field  redefinition $g\rightarrow g'$ satisfying \eqref{FR}. Hence the two actions $S'(g)$ and $S(g')$ are tree-level equivalent.
\vspace{-0.65cm}
\begin{flushright}
$\square$
\end{flushright}
%\vspace{-0.2cm}

As it is obvious from above, in the proof of our theorem it was crucial to use the classical EOM $E_i$. In the theory (\ref{EinsteinBasisFR3}) this implies ${\bf E}=0$ (here no matter source is present).

Now we can explicitly apply the above field redefinition theorem to our class of theories (\ref{EinsteinBasisFR3}), where we do not  include terms with the Riemann tensor $R_{\mu\nu\rho\sigma}$ nor the Weyl tensor $C_{\mu\nu\rho\sigma}$ in the action.
Since we are interested in
$S(g') \equiv S_{{\rm EH}- \Lambda}(g')$ and
$S'(g) \equiv S_{{\rm NL}-\Lambda}(g)$, the relation \eqref{AnselmiC} reads
\be
%S(g') \equiv
%S_{\rm EH}(g') \equiv
 S(g') = S_{{\rm EH}- \Lambda } (g) -  2 \kappa_4^{-2}  \!\int\! d^4 x \sqrt{|g|} \, E_{\mu \nu}(g)  F^{\mu \nu, \rho\sigma}(g)  E_{\rho\sigma}(g)
%\stackrel{=}{gg}
= S'(g)
%\equiv S_{\rm gr}(g)
\,,
\ee
where $E_{\mu\nu}$ is given in (\ref{EHeom}), $F^{\mu \nu, \rho\sigma}(g)$ is defined in (\ref{FEinsteinBasisFR3}), and $\cal V$ compatible with the field redefinition has been introduced in
(\ref{termsVL}).

As a particular implication of the theorem we can always make a field redefinition
to turn the kinetic operator and the propagator for the gravitational fluctuations of the nonlocal theory into the one of Einstein's gravity plus the cosmological constant.
Moreover, when we can properly define asymptotic graviton states in a MSS,
%maximally symmetric spacetime,
all the tree-level on-shell $n$-point functions for the weakly nonlocal theory
(\ref{EinsteinBasisFR3}) are exactly the same as the ones for the Einstein-Hilbert-$\Lambda_{\rm cc}$ gravity (\ref{EHLambda}).

Finally, in view of the theorem proved here, it is clear why at tree-level a class of weakly nonlocal
theories and the local Einstein-Hilbert theory with the cosmological constant
have the same spectrum and the same $n$-point functions, ergo this range of weakly nonlocal
theories is actually local at classical perturbative level.
However, we can not push further the outcome of the theorem because in a theory with an infinite number of derivatives at the moment we do not know the number of nonperturbative degrees of freedom, in contrast to the Einstein-Hilbert theory where the ADM formulation ensures that there are only two degrees of freedom at perturbative and nonperturbative level and around any background. Similarly this theorem likely does not hold at quantum level.

To summarize the content of this section, we proved that the Einstein-Hilbert-$\Lambda_{\rm cc}$ theory (EH-$\Lambda$)  and nonlocal gravity with the presence of cosmological constant term are equivalent at perturbative level. The proof is based on a field redefinition theorem that works perturbatively in $F_{i j}$, but to all perturbative orders in $\Delta_{i j}$. The above result can be therefore seen as a resummation of all perturbative contributions. In the previous sections we have proved that the EOM for both the theories, EH-$\Lambda$ and nonlocal gravity, have the same solutions at the linear order in the gravitational perturbation and, therefore, we inferred that the two theories have the same perturbative spectrum. The theorem in this section guarantees that the classical $n-$point functions (if they can be defined in AdS and/or dS spacetimes) are also the same in the two theories. The theorem is particularly useful in Minkowski spacetime where the $n$-point functions are surely well defined. On the other hand we do not know if the spectrum of the two theories still coincides at nonperturbative level and/or on a general background. Actually, on a general background we expect more degrees of freedom in nonlocal gravity contrary to what happens in the EH-$\Lambda$ theory as recently proved in \cite{CalcagniModestoNardelli1,CalcagniModestoNardelli2}.

At quantum level the two theories are completely different: the EH-$\Lambda$ theory is nonrenormalizable, while nonlocal gravity is finite. Indeed, the field redefinition surely changes the measure in the path-integral and the two theories show  different behaviors at quantum level. We could say that the field redefinition is anomalous because at quantum level other finite operators can be can generated, like for example ${\rm Rieman}^3$, etc., and then the mapping between EH-$\Lambda$ and nonlocal gravity does not work anymore. (In the proof we assumed the action to be quadratic in the EOM, but the finite quantum corrections can violate such assumption.)

\section{More on propagators in weakly nonlocal theories}
\label{revise}

%\subsection{New generalization of AID propagator}

In this section we are going to extend our own construction of propagators in weakly nonlocal theories.
Let us consider a simple example of a weakly nonlocal scalar field theory and its propagator, namely
\begin{equation}
	S=\int d^Dx \, \varphi f(\Box)(\Box+m^2)\varphi \quad \Longrightarrow \quad \Pi=\frac1{f(\Box)(\Box+m^2)} \, .
	\label{simpleaidphi}
\end{equation}
The above theory in most cases comes as a generalization of a local second order theory whose action and propagator respectively read:
\begin{equation}
	S=\int d^Dx \, \varphi  (\Box+m^2)\varphi \quad \Longrightarrow \quad \Pi=\frac1{\Box+m^2}\, .
	\label{simpleaidphilocal}
\end{equation}
One of the most often situations is the wish to constraint $f(\Box)$ such that both theories have the same physical excitations. In this example this means that each theory describes only a single scalar and the mathematical requirement for this is that $f(\Box)$ has no zeros and hence the propagator in (\ref{simpleaidphi}) has no extra new poles on the whole complex plane besides the one at $-m^2$. That is $f(z)\neq 0$ for all $|z|<\infty$, $z \in \mathbb{C}$, where $z=\Box$.

A usual way used to achieve no extra poles in the propagator in (\ref{simpleaidphi}) is to require:
\begin{equation}
	f(z)=e^{\alpha(z)} \quad \text{ where }\alpha(z)\text{ is an entire function} .
	\label{fwas}
\end{equation}
This indeed works and propagates from \cite{Efimov} through all known to us papers on the subject of weakly nonlocal theories. By definition an entire function is a function analytic on the whole complex plane. As such it has no poles in any finite region of the complex plane. The exponent of any finite (and zero) argument is always a nonzero complex number. Actually, the exponent of an entire function is a special entire function with no zeros on the whole complex plane. As a result $f(z)$ in (\ref{fwas}) is always nonzero. Moreover, a particular setup may be required to preserve the normalization of the propagator in the low-energy limit. This implies $f(0)=1$ or equivalently $\alpha(0)=0$. Physically this can be understood as follows: given there is a scale $\Lambda$ which defines the characteristic scale of $f(\Box)$ such that this function is truly $f(\Box/\Lambda^2)$, one may want to see the modified model (\ref{simpleaidphi}) returning to its local counterpart (\ref{simpleaidphilocal}) when $\Lambda\to\infty$. The latter is the local theory limit.

At this point we put the following claim.
\\
{\bf Claim:} The form of $f(\Box)$ given by (\ref{fwas}) is overrestricted and is not necessary as long as the number of degrees of freedom is concerned.
\\
Instead we can prove the following:
\\
{\bf Proposition:} The less restrictive form that is still compatible with the requirements (i) to avoid a generation of new degrees of freedom, (ii) to keep the original normalization in the local limit, and (iii) to preserve and/or improve the UV behavior of the propagator is:
\begin{equation}
	f(z)=\frac{e^{\alpha(z)}}{\beta(z)} \quad \text{ where }\alpha(z),~\beta(z)\text{ are entire functions.}
	\label{fnew}
\end{equation}
Indeed, substituting this in the propagator in (\ref{simpleaidphi}) one gets
\begin{equation}
	\Pi=\frac{\beta(\Box)}{e^{\alpha(\Box)}(\Box+m^2)} \, .
	\label{propnew}
\end{equation}
The exponent in the denominator works exactly as it worked before when the form (\ref{fwas}) was used. The crucial thing to understand is that the new function $\beta(\Box)$ does not change neither of required properties (i)-(iii) as long as it is an entire function. This is a trivial consequence of the definition of an entire function that says that it has no poles on the whole complex plane and as such, our propagator has no new poles as well. The normalization in the local limit can always be preserved by the demand $\beta(0) \exp ({-\alpha(0)})=1$. The UV behavior is subject to a particular choice of the functions $\alpha(\Box)$ and $\beta(\Box)$, which are almost unrestricted so far in any way.

A point of worry is instead the EOM, which we are going to consider in more detail. The EOM can be written as
\begin{equation}
	\frac{e^{\alpha(\Box)}(\Box+m^2)}{\beta(\Box)}\varphi=0 \, .
	\label{eomnew}
\end{equation}
We start with reminding that thanks to the Weierstrass factorization theorem \cite{weierstrass} any entire function $\beta(z)$ can be represented as
\begin{equation}
	\beta(z)=e^{\tilde\beta(z)}\prod_I(z-z_I)^{m_I} \, ,
	\label{weier}
\end{equation}
where $\tilde\beta(z)$ is again an entire function, $z_I$ are roots of $\beta(z)$ and $m_I$ are their multiplicities. First of all, we stress that $\beta(z)$ in the condition of the theorem is an entire function and as such in general $1/\beta(z)$ factor in EOM (\ref{eomnew}) cannot be presented like this. Consequently, and not surprisingly, we do not gain new factors in the numerator of EOM. Having $\alpha(z)$ and $\tilde\beta(z)$ both entire functions we can join them into a redefined function $\tilde\alpha(z)=\alpha(z)-\tilde\beta(z)$. So, without any assumptions we can write the EOM (\ref{eomnew}) as
\begin{equation}
	\frac{e^{\tilde\alpha(\Box)}(\Box+m^2)}{\prod_I(\Box-z_I)^{m_I}}\varphi=0 \, .
	\label{eomnewprod}
\end{equation}
We assume that by construction neither of $z_I$ coincides with $-m^2$. Otherwise, we would immediately write another EOM and propagator. Then a canonical solution originates from the mode
\begin{equation}
	(\Box+m^2)\varphi=0 \, .
	\label{eomnewprodmode}
\end{equation}
Further, it was shown in \cite{mathold} that the exponent operator does not generate new solutions and we can drop it from the consideration of solutions of the EOM. A simple way to see that the denominator does not provide new solutions, which could be associated with new degrees of freedom, is to notice that  we can use the Schwinger and Feynman parametrization to achieve
\begin{equation}
	\begin{split}
	&\frac 1{\prod_I(\Box-z_I)^{m_I}}\varphi=\frac{\Gamma(\sum_I m_I)}{\prod_I\Gamma(m_I)}\int_0^1 \left( \prod_I du_I \right) \frac{\delta(1-\sum_I u_I)\prod_I u_I^{m_I-1}}{ \left[ \sum_I u_I(\Box-z_I) \right]^{\sum_I m_I}}\,
	\varphi= \\
	& \hspace{2.55cm} =\frac{1}{\prod_I\Gamma(m_I)}\int_0^1\left( \prod_I du_Iu_I^{m_I-1} \right) {\delta(1-\sum_I u_I)}\int_0^\infty ds \, s^{\sum_I m_I-1}e^{s\sum_I u_Iz_I}e^{-s\sum_I u_I\Box} \, \varphi
	\, .
\end{split}
	\label{eomnewprodSF}
\end{equation}
This is again an exponential of the d'Alembertian operator acting on the scalar field $\varphi$. Therefore, we can say that no new solutions are generated as long as we can change the order of differentiation and integration. The latter is true as long as a Laplace transform of the scalar field function can be defined. The classical field in turn has to have a well defined Laplace transform in order to be properly quantized.

The things become trickier when we have to define and solve for Green function that is defined as a solution to the fundamental equation
\begin{equation}
	\frac{e^{\tilde\alpha(\Box)}(\Box+m^2)}{\prod_I(\Box-z_I)^{m_I}} \, G(x,x')=\delta(x-x') \, ,
	\label{eomnewprodgreen}
\end{equation}
with appropriate boundary conditions (retarded, advanced, causal, etc.). Here we can act in analogy with the treatment of $1/\Box$ operator in gravity theories like in \cite{woodard2014}. However, a consistent treatment exists for the single inverse d'Alembertian only. We do not need Green functions of this (or any other) kind to proceed, but we very much hope to see this question solved in future works.

Two more comments are in order here. First, the new form (\ref{fnew}) is definitely a significant extension of the class of possible form factors which can enter in weakly nonlocal theories. Second, it will be shown below that such an extension is crucial to guarantee the no-ghost conditions in both regimes: quantum gravity and inflation.

%\subsection{Towards the most general quantum weakly nonlocal theory}

For the case of gravitational theories the propagator (\ref{propnew}), especially the new higher derivative factors must obey several conditions that were first formulated in \cite{Kuzmin, Tombo} and are given above in Sec. \ref{sectionWNL}.
%Namely, we at least require for function $f(z)$:
%\begin{itemize}
%%\renewcommand{\theenumi}%{(\roman{enumi})}

%\item
%the asymptotic behavior of $|f(z)|$ is the same for $z\to\pm\infty$ along the real axis;
%\item
%one can find finite $0<\varphi_\infty<\pi/2$, such that
%$|f(z)| \sim | z |^{\gamma_\infty + \frac D2-1}$  when $|z|\to \infty$, $\gamma_\infty$ is an integer such that $\gamma_\infty\geq[D/2]$ with brackets denoting the integer part of the division, in the conical regions
%$ - \varphi_\infty < {\arg} z < + \varphi_\infty$ and $\pi - \varphi_\infty < {\arg} z < \pi + \varphi_\infty$.
%\end{itemize}
These conditions are aimed at achieving the maximum convergence of loop integrals still preserving the power law falloff of the integrands at infinity. The latter is important to preserve the locality of counterterms and as such to maintain the renormalizability of the theory \cite{modesto,modestoLeslaw,KoshelevStaro}.
Prior the current analysis the conditions in question were considered for the function $f(z)$ as it is given in (\ref{fwas}). However, it is easy to see that no extra complications arise when we have to satisfy the above conditions using the function $f(z)$ in (\ref{fnew}) in kinetic operators for theory not involving gravity or any other non-Abelian gauge theory.

Going further one can easily understand that the form (\ref{fnew}) is again not an ultimate nonlocal factor. That factor was constructed under the assumption that we do not alter the already existing pole at $\Box=-m^2$ in the propagator in (\ref{simpleaidphi}). Under this assumption the nonlocal factor is still maximally general. However, this requirement can in principle be relaxed unless we have some external reasons to maintain this property. Having said this, we understand that we can suggest a function
\begin{equation}
	f(z)=\frac{e^{\alpha(z)}}{\beta(z)}\frac{z+\mu^2}{z+m^2} \quad \text{ where }\alpha(z),~\beta(z)\text{ are entire functions,}
	\label{fnewnew}
\end{equation}
which being substituted in the propagator (\ref{simpleaidphi}) results in
\begin{equation}
	\Pi=\frac{\beta(\Box)}{e^{\alpha(\Box)}(\Box+\mu^2)}\,.
	\label{propnewnew}
\end{equation}
This clearly propagates only a scalar with a new \con mass square $\mu^2$ \con while all other properties remain the same. From here actually no further generalization is seen as long as we preserve the number of poles. The latter property is indeed very much important because new poles will be ghosts due to the Ostrogradsky instability \cite{ostrogradski}.

In an extreme case we can have a very special function
\begin{equation}
	f(z)=\frac{e^{\alpha(z)}}{\beta(z)}\frac1{z+m^2} \quad \text{ where }\alpha(z) ~ {\text{and}} ~\beta(z) ~\text{are entire functions,}
	\label{fnewnew0}
\end{equation}
which being substituted in the propagator (\ref{simpleaidphi}) results in
\begin{equation}
	\Pi=\frac{\beta(\Box)}{e^{\alpha(\Box)}}
	\label{propnewnew0}
\end{equation}
This is a clear analog of a Lagrangian of a $p$-adic theory which has no poles and as such no propagating degrees of freedom in the perturbative vacuum at all.

The following comment follows. Our generalized construction is transparent and obviously valid as long as everything but gravity is concerned. As an immediate example, what we have just discussed, helps in understanding the behavior of quantum perturbations during inflation as will be explained in details in \cite{staroNEW}. This is because the propagator for perturbations explicitly and generically has a numerator factor which is $\beta(\Box)$ above.
Concerning gravity, the propagator (\ref{propnewnew0}) comes together with strongly nonlocal vertices.
%
%however, maybe trickier than expected simply because the metric is involved in the new nonlocal vertices that \cor are not weakly nonlocal and I DO NOT AGREE! If the theory is with weakly nonlocal form factors then the vertices are weakly nonlocal too! \con show up extra poles.
%enters the boxes, which form the essence of our construction.
%The point of worry about is t
The role of these new vertices in the perturbative unitarity is at the moment
%actors will change vertices in a way
not under control and deserves much more investigation.
%Nevertheless, our generalization of the propagator is in accord with the counting of the number of degrees of freedom and this makes it viable for any but gravitational theory. However, the perturbative unitarity and the renormalizability of gravitational theories must be reconsidered, but this is beyond the scope of the present paper.
%%%%%%%%%%%%%%%%%%%%%%%%%%%%%%%%%%%%%%%%%%%%%%%%%%%%%%%%%%%%%%%%%%%%%%%%%%%%%%%%%%%%%%%%%%%%%

\section{Conclusions}
In this paper we explicitly proved that all the weakly nonlocal gravitational theories consistent at quantum level have exactly the same classical properties as Einstein's gravity at linear level when studied perturbatively around any maximally symmetric spacetime background. These theories differ only for the presence or not of the Weyl tensor in the nonlocal operators quadratic in curvatures, but the outcome is always the same. Namely, the quadratic action at the second order in the graviton perturbations around any MSS can be recast in the form of the Einstein-Hilbert quadratized action up to exponential form factors in front of the corresponding projectors for the spin-two and spin-zero components. For one out of the two ranges of theories, namely the one without Weyl or Riemann tensor in the action, we proved, making use of a field redefinition theorem, that the theory is perturbatively (in the entire function defining the field redefinition) equivalent to the Einstein-Hilbert action in the presence of a cosmological constant. This statement holds to all orders in Taylor expansion in the field redefinition of the metric tensor. Moreover, the field redefinition theorem, when the graviton's asymptotic states on a MSS are properly defined, endorses that all tree-level $n$-point scattering amplitudes in the weakly nonlocal theory coincide with the ones of Einstein-Hilbert gravity with cosmological constant on the same MSS background.

At quantum level, for one out of the two classes of theories (namely the one in Weyl's basis) we explicitly proved that all the beta functions can be made to vanish. Therefore, the quantum theory is finite (in DIMREG scheme) on any MSS. Certainly, also the theory in the Ricci basis enjoys the same convergence properties.

We can finally claim that the weakly nonlocal theories are perturbatively well defined, unitary (as long as the Einstein-Hilbert is), and finite at quantum level on any maximally symmetric space. Having an ultraviolet complete theory for gravity in the quantum field theory framework, we can now study the implications and/or applications in the AdS/CFT domain.
 The AdS/CFT correspondence is clearly defined, but there is no clear definition of the dS/CFT correspondence, unless one appeals to a nonlocal map between the AdS and dS spaces as discussed in
\cite{Balasubramanian:2002zh}. However, we remark that our construction is valid and works equally well for both signs of the cosmological constant. Firstly, preliminary results show that the transition from dS to AdS can be reached as an effect of RG flow of the couplings of the theory. Secondly, of course, we understand that for holding of gauge/gravity duality even simple kinematical conditions must be satisfied (like the equivalence between the group of isometries of AdS and the conformal group in flat Minkowski spacetime) and these are not true on dS. However, here we treat dS/AdS as backgrounds and for the the moment quantum  backreaction is neglected. We may express a belief that including backreaction effects in a particular class of theories studied in this paper will show some preference towards the sign of the curvature of the background and then we could ultimately decide whether we can or why we cannot extend the duality to a dS spacetime.  We are interested to export all the results obtained in string theory and in the AdS/CFT correspondence to nonlocal quantum gravity, and we will surely invest time and resources on this topic in the next future.  We believe it will be interesting to see whether nonlocal gravity could shed light on various conceptual problems associated with possible dS/CFT correspondence.

%by Balasubramanian et al, in Class.\ Quant.\ Grav.\ {\bf 19}, 5655 (2002) [Annals Phys.\ {\bf 303}, 59 (2003)] [hep-th/0207245])In particular, the nonlocal nature of the theory could shed light on various conceptual problems associated with a good definition of the dS/CFT correspondence as pointed out in Class.\ Quant.\ Grav.\ {\bf 19}, 5655 (2002) [Annals Phys.\ {\bf 303}, 59 (2003)] [hep-th/0207245]).

%\subsection{Power counting}
%{\color{red} The power counting analysis can be done around flat spacetime even though the Minkowski vacuum is not an exact solution. } Indeed what do matter are the higher derivative operators in the ultraviolet regime.
%Using the Batalin-Vilkovisky formalism Piva and Anselmi have recently shown that the superrenormalizability
%of the theory is %guaranteed in any background, therefore we have background independence.

\acknowledgments
%
%AK is supported in part by FCT Portugal investigator project IF/01607/2015 and by the grant UID/MAT/00212/2013. KSK acknowledges the support from the FCT grant SFRH/BD/51980/2012.
%
A.K. is supported by FCT Portugal investigator project IF/01607/2015, FCT Portugal fellowship  SFRH/ BPD/ 105212/2014, and in part by FCT Portugal
grant  UID/MAT/00212/2013.

\section*{Appendix: Variations}
Here we collect the results about variations on a maximally symmetric background of  operators quadratic in curvature. We write them in a manifestly self-adjoint form. First the variation of action written with Weyl tensors~reads,
%
%{\color{red} ATTENTION: In this section $\Lambda_{\rm cc}$ is the cosmological constant and not the nonlocality scale}.
{%\small{
\be
&&\frac{1}{2}\delta^{2}\left(\int\!d^{D}x\sqrt{|g|}C_{\mu\nu\rho\sigma}{\cal F}\left(\square\right)C^{\mu\nu\rho\sigma}\right)\underset{\mathrm{MSS}}{=} \nonumber
\\
&&
=\int\!d^{d}x\sqrt{|g|}\left[h_{\mu\nu}\left(\frac{2D(D-3)}{(D-2)(D-1)^{2}}\Lambda_{\rm cc}^{2}-\frac{(D-3)(D+2)}{(D-2)(D-1)}\Lambda_{\rm cc}\square+\frac{D-3}{D-2}\square^{2}\right){\cal F}\left(\square+2\frac{D-2}{D-1}\Lambda_{\rm cc}\right)h^{\mu\nu}\right. \nonumber  \\
&&
+h\left[\left(-\frac{2(D-3)}{(D-2)(D-1)^{2}}\Lambda_{\rm cc}^{2}+\frac{(D-3)(D+2)}{D(D-2)(D-1)}\Lambda_{\rm cc}\square-\frac{D-3}{D(D-2)}\square^{2}\right){\cal F}\left(\square+2\frac{D-2}{D-1}\Lambda_{\rm cc}\right) \right.\nonumber   \\
&& \left.
-\left(\frac{2\left(D-3\right)}{D(D-2)(D-1)}\Lambda_{\rm cc}\square+\frac{D-3}{D(D-2)(D-1)}\square^{2}\right){\cal F}\left(\square+4\Lambda_{\rm cc}\right)\right]h \nonumber
\\
&&
+h_{\mu\nu}\nabla^{\mu}\nabla^{\nu}\left(\frac{2(D-3)}{(D-2)(D-1)}\Lambda_{\rm cc}+\frac{D-3}{(D-2)(D-1)}\square\right){\cal F}\left(\square+4\Lambda_{\rm cc}\right)h  \nonumber \\
&&
+ h\left(\frac{2\left(D-3\right)}{(D-2)(D-1)}\Lambda_{\rm cc}+\frac{D-3}{(D-2)(D-1)}\square\right){\cal F}\left(\square+4\Lambda_{\rm cc}\right)\nabla_{\mu}\nabla_{\nu}h^{\mu\nu} \nonumber
\\
&&
\left.+h_{\mu\nu}\nabla^{\mu}\left(-\frac{2(D-3)}{(D-2)(D-1)}\Lambda_{\rm cc}-\frac{2(D-3)}{D-2}\square\right){\cal F}\left(\square+3\Lambda_{\rm cc}\right)\nabla_{\rho}h^{\nu\rho}
\right. \nonumber \\&& \left.
+\frac{D-3}{D-1}h_{\mu\nu}\nabla^{\mu}\nabla^{\nu}{\cal F}\left(\square+4\Lambda_{\rm cc}\right)\nabla_{\rho}\nabla_{\sigma}h^{\rho\sigma}\right] \, .
\ee
}
\hspace{-0.4cm}
Next we introduce the definition $\hat{R} = R - D\Lambda_{\rm cc}=R-\overline{R}$ and we evaluate the following variation,
\be
&& \hspace{-1.5cm}
\frac{1}{2}\delta^{2}\left(\int\!d^{D}x\sqrt{|g|}\hat{R}{\cal F}\left(\square\right)\hat{R}\right)\underset{\mathrm{MSS}}{=} %\nonumber \\&& %\hspace{-1cm}=
\int\!d^{D}x\sqrt{|g|}\Big[h\left(\Lambda_{\rm cc}+\square\right)^{2}{\cal F}\left(\square\right)h-h_{\mu\nu}\nabla^{\mu}\nabla^{\nu}\left(\Lambda_{\rm cc}+\square\right){\cal F}\left(\square\right)h
\nonumber  \\
&& \hspace{4.5cm}
-h\left(\Lambda_{\rm cc}+\square\right){\cal F}\left(\square\right)\nabla_{\mu}\nabla_{\nu}h^{\mu\nu} %
+h_{\mu\nu}\nabla^{\mu}\nabla^{\nu}{\cal F}\left(\square\right)\nabla_{\rho}\nabla_{\sigma}h^{\rho\sigma}\Big]
\, .
\ee
On a MSS we have the following values of the background curvatures,
\be
\overline{R}_{\mu\nu\rho\sigma}=\frac{2\Lambda_{\rm cc}}{D-1}g_{\mu[\rho}g_{\sigma]\nu} \, , \quad\overline{R}_{\mu\nu}=\Lambda_{\rm cc} g_{\mu\nu} \, ,
\quad \overline{R}=D \Lambda_{\rm cc} \, .
\ee
%Other u
An useful formula is:
\be
\hat{R}{\cal F}\left(\square\right)\hat{R}=R{\cal F}\left(\square\right)R-2 D {\cal F}_{0}\Lambda_{\rm cc} R + D^{2}{\cal F}_{0}\Lambda_{\rm cc}^{2} \, , \quad
\mbox{ where}\quad
 {\cal F}_{0}={\cal F}\left(0\right).
 \ee
The variations of the cosmological constant $a_\Lambda$ and the Einstein-Hilbert actions read as follows \cite{Leslawthesis},
\be
&& \frac{1}{2}\delta^{2}\left(\int\! d^{D}x\sqrt{|g|} a_\Lambda \right)=\int\!d^{D}x\sqrt{|g|}\left(\frac{1}{8}h^{2}-\frac{1}{4}h_{\mu\nu}h^{\mu\nu}\right) a_\Lambda \, , \\
&& \frac{1}{2}\delta^{2}\left(\int\!d^{D}x\sqrt{|g|}R\right)\underset{\mathrm{MSS}}{=}
%\nonumber \\&&
\int\!d^{D}x\sqrt{|g|}\left[ \frac{1}{4}h_{\mu\nu}\square h^{\mu\nu}-\frac{1}{4}h\square h+\frac{1}{4}h\nabla_{\mu}\nabla_{\nu}h^{\mu\nu}+\frac{1}{4}h_{\mu\nu}\nabla^{\mu}\nabla^{\nu}h
\right. \nonumber \\
&& \hspace{3.5cm} \left. -\frac{1}{2}h_{\mu}{}_{\nu}\nabla^{\mu}\nabla_{\rho}h^{\nu\rho}
+\Lambda_{\rm cc}\left(-\frac{D^{2}-3 D +4}{4 \left( D-1\right)}h_{\mu\nu}h^{\mu\nu}+\frac{D^{2}-5 D +8}{8\left(D-1\right)}h^{2}\right) \right]  \, .
\ee
%%%%%%
%%%%%%
For completeness we give also the expression for the variation (in normal form) in $D=4$,
\be
\label{varffd4}
&& \frac{1}{2}\delta^{2}\left(\int\!d^{4}x\sqrt{|g|}C_{\mu\nu\rho\sigma}{\cal F}\left(\square\right)C^{\mu\nu\rho\sigma}\right)\underset{\mathrm{MSS}}{=}\int\!d^{4}x\sqrt{|g|}\left[h_{\mu\nu}\left(\frac{4}{9}\Lambda_{\rm cc}^{2}-\Lambda_{\rm cc}\square+\frac{1}{2}\square^{2}\right){\cal F}_{2}h^{\mu\nu}+\right. \nonumber  \\
&&
+h\left(-\frac{1}{9}\Lambda_{\rm cc}^{2}+\frac{5}{18}\Lambda_{\rm cc}\square-\frac{1}{6}\square^{2}\right){\cal F}_{2}h
%&&
+h_{\mu\nu}\nabla^{\mu}\nabla^{\nu}\left(-\frac{1}{3}\Lambda_{\rm cc}+\frac{1}{6}\square\right){\cal F}_{2}h+h\nabla_{\mu}\nabla_{\nu}\left(-\frac{1}{9}\Lambda_{\rm cc}+\frac{1}{6}\square\right){\cal F}_{2}h^{\mu\nu}
\nonumber \\
&&
\left.+h_{\mu\nu}\nabla^{\mu}\nabla_{\rho}\left(\frac{4}{3}\Lambda_{\rm cc}-\square\right){\cal F}_{2}h^{\nu\rho}+\frac{1}{3}h_{\mu\nu}\nabla^{\mu}\nabla^{\nu}\nabla_{\rho}\nabla_{\sigma}{\cal F}_{2}h^{\rho\sigma}\right],
\ee
where we introduced a short notation for the form factor with a translated argument,
\be
{\cal F}_{2} \equiv {\cal F}\left(\square+\frac{4}{3}\Lambda_{\rm cc}\right) \, .
\ee
We observe an interesting fact, that all dependence on form factor in \eqref{varffd4} is only via shifted one ${\cal F}_{2}$.
The variation of the nonlocal Weyl square operator on a MSS background can be written also in a manifestly self-adjoint form in $D=4$, namely
\be
&&\frac{1}{2}\delta^{2}\left(\int\!d^{4}x\sqrt{|g|}C_{\mu\nu\rho\sigma}{\cal F}\left(\square\right)C^{\mu\nu\rho\sigma}\right)\underset{\mathrm{MSS}}{=}\nonumber
\\
&&
=\int\!d^{4}x\sqrt{|g|}\left[h_{\mu\nu}\left(\frac{4}{9}\Lambda_{\rm cc}^{2}-\Lambda_{\rm cc}\square+\frac{1}{2}\square^{2}\right){\cal F}\left(\square+\frac{4}{3}\Lambda_{\rm cc}\right)h_{\mu\nu}+\right.\nonumber
\\
&&
+h\left[\left(-\frac{1}{9}\Lambda_{\rm cc}^{2}+\frac{1}{4}\Lambda_{\rm cc}\square-\frac{1}{8}\square^{2}\right){\cal F}\left(\square+\frac{4}{3}\Lambda_{\rm cc}\right)-\left(\frac{1}{12}\Lambda_{\rm cc}\square+\frac{1}{24}\square^{2}\right){\cal F}\left(\square+4\Lambda_{\rm cc}\right)\right]h+\nonumber
\\
&&
+h_{\mu\nu}\nabla_{\mu}\nabla_{\nu}\left(\frac{1}{3}\Lambda_{\rm cc}+\frac{1}{6}\square\right){\cal F}\left(\square+4\Lambda_{\rm cc}\right)h+h\left(\frac{1}{3}\Lambda_{\rm cc}+\frac{1}{6}\square\right){\cal F}\left(\square+4\Lambda_{\rm cc}\right)\nabla_{\mu}\nabla_{\nu}h_{\mu\nu}-\nonumber
\\
&&
\left.-h_{\mu\nu}\nabla_{\mu}\left(\frac{1}{3}\Lambda_{\rm cc}+\square\right){\cal F}\left(\square+3\Lambda_{\rm cc}\right)\nabla_{\rho}h_{\nu\rho}+\frac{1}{3}h_{\mu\nu}\nabla_{\mu}\nabla_{\nu}{\cal F}\left(\square+4\Lambda_{\rm cc}\right)\nabla_{\rho}\nabla_{\sigma}h_{\rho\sigma}\right] \, .
\ee

It can be easily seen that in this self-adjoint form we encounter 3 different shifts of the argument of the form factor by $4/3\Lambda_{\rm cc}$, $3\Lambda_{\rm cc}$ and $4\Lambda_{\rm cc}$ respectively.

The above results for the second variations were checked using various methods. First, all expressions can be put in the self-adjoint form of the operator of the second order variational derivative. Second, all the variations, except for the term with cosmological constant only, are invariant under the substitution $h_{\mu\nu}\to h_{\mu\nu}+\nabla_{(\mu}\xi_{\nu)}$ for all left or right instances of the fluctuations of metric, where $\xi_\nu$ is an arbitrary vector field and when the on-shell background is used. This is the statement of gauge-invariance of the action with respect to general coordinate transformations. Last but not less important, the second variations were checked against conformal invariance of the action with two Weyl tensors in $D=4$. More precisely, it had been checked that for metric fluctuations of the form $h_{\mu\nu}= \omega^2(x)g_{\mu\nu}$ the second variation of such action on any MSS background vanishes.

\end{document}